\documentclass[a4paper,12pt]{article}
\pdfoutput=1

\usepackage{color}
\usepackage{multirow}
\usepackage{longtable}
\usepackage{xspace}
\usepackage{listings}
\usepackage{changepage}
\usepackage{amsmath,amssymb}
\usepackage{slashed}
\usepackage{xcolor,colortbl}
\usepackage{hyperref}
\usepackage{adjustbox}
\usepackage{textcomp}

\RequirePackage[T1]{fontenc}
\RequirePackage{graphicx}
\RequirePackage[numbers,sort&compress]{natbib}

\hypersetup{colorlinks,bookmarksopen,bookmarksnumbered,linkcolor=blus,pdfstartview=FitH,urlcolor=blus,citecolor=verde}

\definecolor{tit}{rgb}{0.1,0.2,0.4}
\definecolor{blus}{cmyk}{1,1,0,0.6}
\definecolor{verde}{cmyk}{0.92,0,0.59,0.25}
\definecolor{lblue}{rgb}{0.7,0.7,1}

\newcommand{\Dlr}{\overset\leftrightarrow{D}}
\newcommand{\DlrImu}{\Dlr_\mu \hspace*{-0.16cm}{}^I}

\newcommand{\vp}{\varphi}

\newcommand{\Op}{\mathcal{O}}
\newcommand{\hc}{\mathrm{h.c.}}

\newcommand{\nn}{\nonumber}

\newcommand{\myv}[1]{{{\rule{0cm}{0.35cm}#1}}}

\newcommand{\eq}[1]{\begin{equation} #1 \end{equation}}
\newcommand{\eqa}[1]{\begin{eqnarray} #1 \end{eqnarray}}
\newcommand{\Reff}[1]{Ref.\,\cite{#1}}
\newcommand{\Sec}[1]{Section\,\ref{#1}}
\newcommand{\App}[1]{Appendix\,\ref{#1}}

\newcommand{\rut}[1]{{\tt \color{verde} #1}}
\newcommand{\gev}{~\text{GeV}}
\newcommand{\tev}{~\text{TeV}}

\definecolor{shaded}{RGB}{245,245,245}

\lstdefinestyle{mathematica}{
  basicstyle=\ttfamily\mdseries,
  backgroundcolor=\color{shaded},
  language=Mathematica,
  frame=false	
}

\lstdefinestyle{WCsInput}{
  basicstyle=\ttfamily\mdseries,
  language=bash,
  frame=single,
  title=\hspace{13.8cm}{\tt WCsInput.dat}	
}

\lstdefinestyle{WCsInput-json}{
  basicstyle=\ttfamily\mdseries,
  language=bash,
  frame=single,
  title=\hspace{13.3cm}{\tt WCsInput.json}	
}

\lstdefinestyle{WCsInput-yaml}{
  basicstyle=\ttfamily\mdseries,
  language=bash,
  frame=single,
  title=\hspace{13.3cm}{\tt WCsInput.yaml}	
}

\lstdefinestyle{SMInput}{
  basicstyle=\ttfamily\mdseries,
  language=bash,
  frame=single,
  title=\hspace{14cm}{\tt SMInput.dat}	
}

\lstdefinestyle{Options}{
  basicstyle=\ttfamily\mdseries,
  language=bash,
  frame=single,
  title=\hspace{14cm}{\tt Options.dat}	
}

\definecolor{Gray}{gray}{0.95}
\definecolor{RGray}{gray}{0.85}
\definecolor{CGray}{gray}{0.92}
\definecolor{dgray}{gray}{0.4}

\definecolor{color1}{rgb}{0.9,.4,.2}
\definecolor{color2}{rgb}{0.3,.6,.7}
\definecolor{color3}{rgb}{0.7,.2,.7}

\newcommand{\pkg}[1]{{\tt #1}\xspace}
\newcommand{\dsix}{\pkg{DsixTools}}
\newcommand{\dsixv}[1]{\pkg{DsixTools\,$#1$}}
\newcommand{\dsixbf}{\pkg{\bf DsixTools}}

\newcommand{\mathe}{\pkg{Mathematica}}
\newcommand{\json}{\pkg{JSON}}
\newcommand{\yaml}{\pkg{YAML}}
\newcommand{\real}[1]{{\color{red} #1}}

\allowdisplaybreaks

\begin{document}

\hspace{-4mm}
\begin{minipage}{16cm}

\vspace{-0.7cm}

\begin{flushright}
{\small
MITP/20-061 \\
IFIC/20-50
}
\end{flushright}

\vspace{-1.3cm}
\noindent\includegraphics[width=4cm]{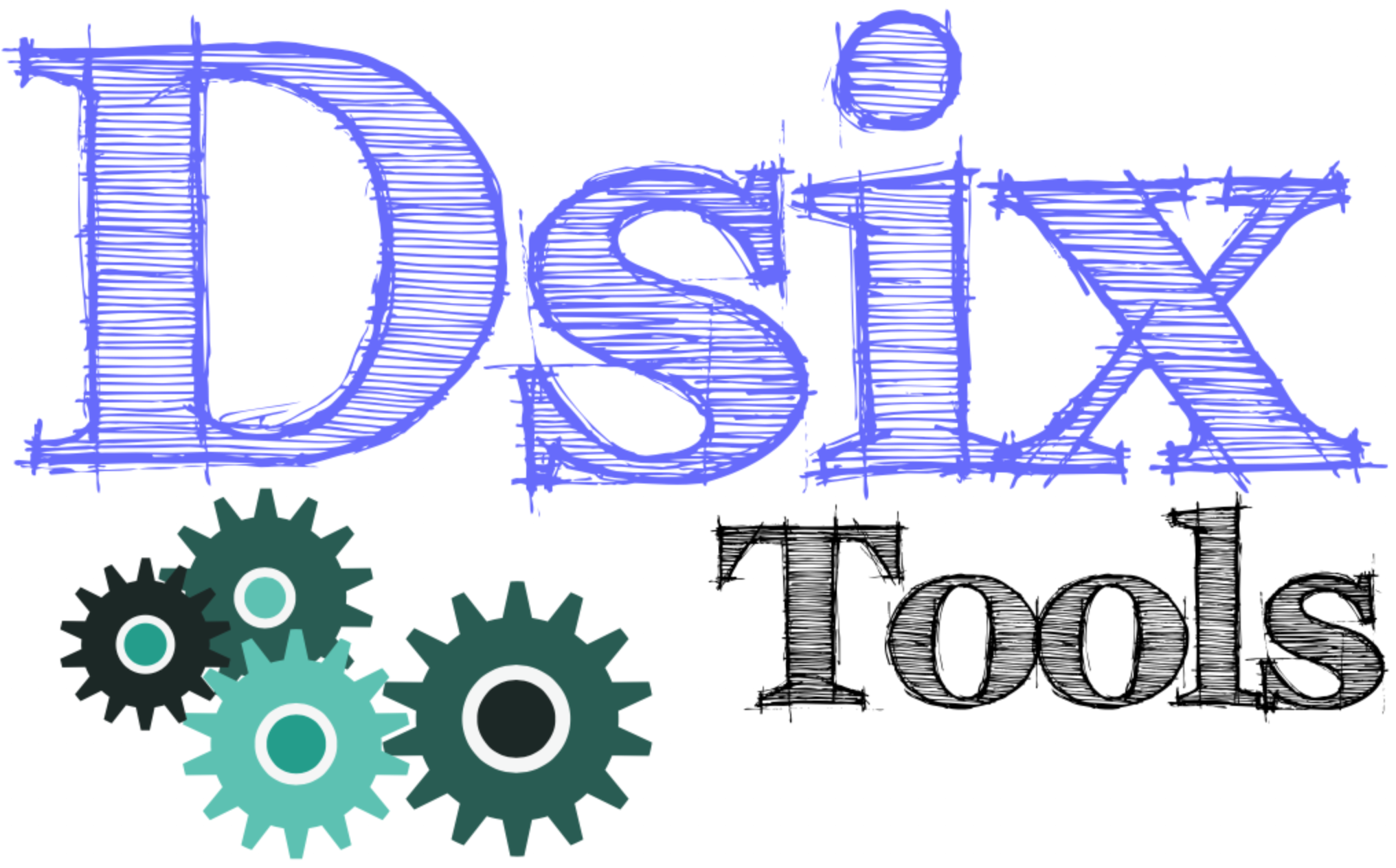}

\vspace{8mm}

\hrule
\vspace{2mm}
\begin{flushleft}
{\bf 
\huge{\color{dgray} DsixTools 2.0}\\[3mm]
\vspace*{0.3cm}
\LARGE{The Effective Field Theory Toolkit}}\\[5mm]
\hrule

\vspace{7mm}

{
\large
Javier Fuentes-Mart\'in$^{a}$,
\\[1.4mm]
Pedro Ruiz-Femen\'ia$^{b}$,
\\[1.1mm]
Avelino Vicente$^{c}$,
\\[2.2mm]
Javier Virto$^{d}$
}\\[5mm]
{\small
{\it $^a$ 
PRISMA+ Cluster of Excellence \& Mainz Institute for Theoretical Physics,\\
\hspace{2.1mm} Johannes Gutenberg University, 55099 Mainz, Germany,}\\[2mm]
{\it $^b$ 
Departament de Matem\`atiques per a l'Economia i l'Empresa,\\
\hspace{2.1mm} Universitat de Val\`encia, E-46022 Val\`encia, Spain}\\[2mm]
{\it $^c$ Instituto de F\'{\i}sica Corpuscular and Departament de F\'{\i}sica Te\`{o}rica,\\
\hspace{3.0mm}Universitat de Val\`encia - CSIC, E-46071 Val\`encia, Spain}\\[2mm]
{\it $^d$ Departament de F\'isica Qu\`antica i Astrof\'isica and ICCUB,\\
\hspace{1.5mm} Universitat de Barcelona, 08028 Barcelona, Catalunya}
}
\end{flushleft}

\vspace{6mm}

\linespread{1.05}

\noindent  {\large\bf Abstract}\\[2mm]
\dsix is a \mathe package for the handling of the Standard Model Effective Field Theory (SMEFT) and the Low-energy Effective Field Theory (LEFT) with operators up to dimension six, both at the algebraic and numerical level.
\dsix contains a visually accessible and operationally convenient repository of all operators and parameters of the SMEFT and the LEFT. 
This repository also provides
information concerning symmetry categories and number of degrees of freedom, and routines that allow to implement this information on global expressions (such as decay amplitudes and cross-sections).
\dsix also performs weak basis transformations, and implements the full one-loop Renormalization Group Evolution in both EFTs (with SM beta functions up to five loops in QCD), and the full one-loop SMEFT-LEFT matching at the electroweak scale.

\vspace{10pt}

\end{minipage}

\thispagestyle{empty}

\begin{quote}
{\large\noindent
}

\numberwithin{equation}{section}

\end{quote}

\tableofcontents

\newpage

\linespread{1.17}

\setcounter{footnote}{0}

\section{Introduction}
\label{sec:intro}

The experimental success of the Standard Model (SM) of particle physics and the absence of new physics (NP) signals  after LHC run 2,  seem to indicate the presence of a mass gap between the Electroweak (EW) scale and the scale of potential new dynamics. If this is the case, non-standard effects in processes at energy scales much smaller than the scale of NP can be described within Effective Field Theory (EFT).

Above the EW scale, the relevant EFT which contains the SM as the low-energy limit is called the Standard Model EFT (SMEFT). The SMEFT accounts for the effect of unknown heavy degrees of freedom by extending the SM Lagrangian with higher-dimensional operators invariant under the SM gauge group. The dominant NP contributions to most of the processes of phenomenological interest are then parametrized by Wilson Coefficients (WCs) of SMEFT operators of canonical dimension five and six~\cite{Buchmuller:1985jz}.

Below the EW scale, heavy SM particles (massive bosons and the top quark) also decouple, and the dynamics is described by the Low-Energy EFT (LEFT). This theory consists of the QCD and QED Lagrangians for the light SM fermions complemented with a set of higher-dimensional operators compatible with the gauge symmetries of QED and QCD. The Wilson coefficients of these higher dimensional operators encode all the physics related to heavy SM states and the NP degrees of freedom, dominated again by operators of canonical dimension five and six~\cite{Jenkins:2017jig}. The LEFT is more general than the SMEFT since it is still the correct low-energy EFT when there are new particles at the EW scale. However, under the SMEFT hypothesis, one can define the LEFT (fix its WCs) by doing a matching to the SMEFT at the EW scale.

The basis for automation of calculations within these two EFTs arises from work done within the last decade. First, a complete non-redundant operator basis for the SMEFT up to dimension six was derived in~\Reff{Grzadkowski:2010es} (aka the \textit{Warsaw basis}). The complete set of one-loop anomalous dimensions of the operators in the Warsaw basis was then calculated in a series of papers~\cite{Jenkins:2013zja,Jenkins:2013wua,Alonso:2013hga,Alonso:2014zka}. Similarly, a complete and non-redundant basis for the LEFT up to dimension six was constructed in~\Reff{Jenkins:2017jig} (aka the \textit{San Diego basis}), and the full one-loop anomalous dimensions were calculated in~\Reff{Jenkins:2017dyc}.
Finally, the tree-level and one-loop matching between the LEFT and the SMEFT was performed in~Refs.~\cite{Jenkins:2017jig} and~\cite{Dekens:2019ept}, respectively.

These advances, together with simultaneous theoretical developments occurring in the field (such as the matching of specific models to the SMEFT at one loop~\cite{Henning:2014wua,Drozd:2015rsp,delAguila:2016zcb,Boggia:2016asg,Henning:2016lyp,Ellis:2016enq,Fuentes-Martin:2016uol,Zhang:2016pja,Ellis:2017jns,Kramer:2019fwz,Ellis:2020ivx}, or the automation of calculations by means of several computer tools~\cite{Brivio:2019irc,Celis:2017hod,Gripaios:2018zrz,Criado:2019ugp,Dedes:2019uzs,Criado:2017khh,Bakshi:2018ics,Aebischer:2018bkb,Hartland:2019bjb,Aebischer:2018iyb,EOS,Straub:2018kue,Brivio:2017btx}), pave the way to the systematic use of EFT methods in the analysis of new physics models. The power of the this approach is that it allows to relate physics at disparate energy scales, in our case properties of the high-energy dynamics at the new physics scale~$\Lambda_{\rm UV}$, with measurements that take place at low energies, while performing an expansion in $1/\Lambda_{\rm UV}$ that allows to keep leading new physics
effects in a consistent manner.

The \mathe\footnote{\mathe~is a product from Wolfram Research, Inc.~\cite{wolfram}.}~package \dsix~\cite{Celis:2017hod} was developed as a tool to implement such automated calculations.
Since the first release of~\dsix in 2017, further development of the package has occurred in two directions: 1) implementation of new theory results (such as moving from the WET~\cite{Aebischer:2017gaw} to the LEFT, and the implementation of higher-order effects), and 2) improvements and refinements at the front-end and operational levels (new routines, input, documentation, faster methods for RG evolution, and notation).
The result of these developments is the new release \dsixv{2.0}, which is available at
\begin{center}
\href{https://dsixtools.github.io}{https://dsixtools.github.io}
\end{center}
This paper presents a description of the program and its new features.


\section{DsixTools in a nutshell}
\label{sec:nutshell}

\subsection{Overview of \texorpdfstring{\dsixv{2.0}}{DsixTools 2.0}}

\dsix is a \mathe package for analytical and numerical computations within the SMEFT and the LEFT. 
It features routines devoted to RGE running (in the SMEFT and in the LEFT), matching between the two theories, basis transformation, input reading (with consistency checks) and output generation. 
\dsix also contains a comprehensive and pedagogical repository with routines that allow the user to display lists of operators with certain properties, and information on WCs in the SMEFT and the LEFT.

The current version of \dsix (\dsixv{2.0}) fully implements the one-loop SMEFT RGEs, the complete one-loop matching between the SMEFT and the LEFT, and the one-loop LEFT RGEs, all up to operators of canonical dimension six.
In what concerns the SMEFT RGE running, \dsix contains:

\begin{itemize}
\item Three-loop SM RGEs from Refs.~\cite{Bednyakov:2012rb,Bednyakov:2012en,Bednyakov:2013eba,Bednyakov:2014pia}, as well as five-loop QCD corrections to the running of the strong gauge coupling and quark Yukawa couplings from Refs.~\cite{vanRitbergen:1997va,Vermaseren:1997fq,Baikov:2017ujl}.\,\footnote{
The one- and two-loop SM RGEs were computed in~\cite{Machacek:1983tz,Machacek:1983fi,Machacek:1984zw} and \cite{Luo:2002ey}, respectively.
}

\item One-loop RGEs for the dimension-six operators in the Warsaw basis from Refs.~\cite{Jenkins:2013zja,Jenkins:2013wua,Alonso:2013hga}.\footnote{
We have taken into account the errata published in~\href{http://einstein.ucsd.edu/smeft/}{http://einstein.ucsd.edu/smeft/}.
}

\item One-loop RGEs for the dimension-six baryon-number-violating operators
  from~\Reff{Alonso:2014zka}. 

\item One-loop RGE for the dimension-five lepton-number-violating operator from~\Reff{Antusch:2001ck}. 
\end{itemize} 

\vspace{0.1cm}

\noindent Regarding the SMEFT-LEFT matching, \dsix implements:

\begin{itemize}
\item The tree-level matching of the SMEFT Warsaw basis to the LEFT San Diego basis at the electroweak scale, using the results of~\Reff{Jenkins:2017jig}. We have independently derived the matching relations (in two different ways), finding full agreement.

\item The complete one-loop matching of the SMEFT Warsaw basis to the LEFT San Diego basis at the electroweak scale, using the results of~\Reff{Dekens:2019ept}.
\end{itemize}

\vspace{0.1cm}

\noindent Finally, \dsix also implements several results related to the RGE running in the LEFT:

\begin{itemize}
\item Four-loop QCD corrections to the strong coupling beta function and quark mass anomalous dimensions from~\Reff{Chetyrkin:2000yt}.

\item One-loop RGEs for all LEFT operators up to dimension six in the San Diego basis from~\Reff{Jenkins:2017dyc}.

\end{itemize}

\begin{figure}
\centering
\includegraphics[width=8cm]{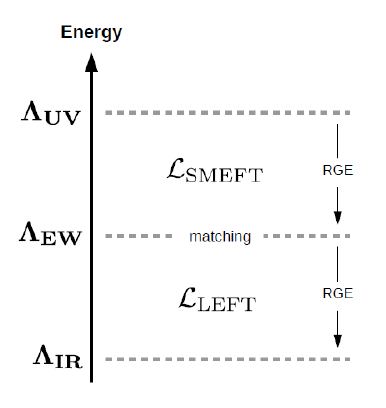}
\caption{\small Scketch of the \dsix matching-running routine. The \dsix terminology is: $\Lambda_\text{UV}$=\rut{HIGHSCALE}, $\Lambda_\text{EW}$=\rut{EWSCALE} and $\Lambda_\text{IR}=\rut{LOWSCALE}$. The default is \rut{EWSCALE}=$M_Z=91.1876\gev$.
}
\label{fig:diagram}
\end{figure}

The structure of \dsix is illustrated in Fig.~\ref{fig:diagram}, where one can also see how they relate to the different energy ranges and effective theories. Relevant details of the SMEFT and LEFT implementations are given in Appendices \ref{sec:SMEFT}--\ref{ap:parameters}, where our conventions are also presented.

\subsection{Differences with \texorpdfstring{\dsixv{1.0}}{DsixTools 1.0}}

The list of improvements and changes that features the new version with respect to the original version published in 2017 is substantial, and programs written with \dsixv{1.0} will most likely not work with \dsixv{2.0}. Thus we collect here a summary of the most relevant changes:

\begin{itemize}

\item \dsixv{2.0} is now very easy to install, directly within \mathe.
See~\Sec{sec:down}.

\item The notation for WCs has changed such that now they are dimensionful.
For example the SMEFT Lagrangian is given by:
\begin{equation}
\mathcal{L}_{\rm SMEFT} = \mathcal{L}_{\rm SM}^{(4)} + \sum_k C_k^{(5)} Q_k^{(5)}  + \sum_k C_k^{(6)} Q_k^{(6)} + \mathcal{O}\left( \frac{1}{\Lambda_{\rm UV}^3} \right) \, ,
\end{equation}
with $C_k^{(5)}\sim \Lambda_{\rm UV}^{-1}$ and $C_k^{(6)}\sim \Lambda_{\rm UV}^{-2}$.
Same principle applies also to the LEFT WCs.

\item The WET~\cite{Aebischer:2017gaw} basis has been superseded by the LEFT, in order to implement all the new results derived within the latter.

\item Nomenclature for operators and Wilson coefficients has been modified, mainly for global convenience and consistency, and in part to make it closer to more common standards (\emph{e.g.} WCxf~\cite{Aebischer:2017ugx} or FeynRules~\cite{Dedes:2019uzs}).

First, all operators in the SMEFT start with {\tt Q} (\emph{e.g.} $Q_{\phi \ell}^{(3)}=\rut{QHl3}$) while the ones in the LEFT start with {\tt O} (\emph{e.g.} $\Op_{ud}^{(V8,LL)}=\rut{OudV8LL}$)

Second, Wilson coefficients in the SMEFT start with {\tt C} (\emph{e.g.} $[C_{\phi \ell}^{(3)}]_{12}=\rut{CHl3[1,2]}$) while the ones in the LEFT start with {\tt L} (\emph{e.g.} $[L_{ud}^{(V8,LL)}]_{1213}=\rut{LudV8LL[1,2,1,3]}$).
In \dsixv{1.0}, flavor matrices were specified as {\tt WC[$name$]}, where $name$ was not the same as the name of the Wilson coefficient (\emph{e.g.} {\tt WC[$\varphi$l3]} vs. {\tt $\varphi$L3[1,2]}).
Flavor matrices in \dsixv{2.0} have the same name as the WCs but with an `{\tt M}' in front, \emph{e.g.}
\eqa{
\rut{MCHl3} &=& \{ \{[C_{\phi \ell}^{(3)}]_{1,1}, [C_{\phi \ell}^{(3)}]_{1,2}, [C_{\phi \ell}^{(3)}]_{1,3} \},\cdots \} \ ,
\nonumber\\[2mm]
\rut{MLudV8LL}&=& \{\{\{\{[L_{ud}^{(V8,LL)}]_{1111},[L_{ud}^{(V8,LL)}]_{1112},\cdots\},\cdots\}\}\} \ .
\nonumber
}
In addition, characters that are not trivially easy to type in \mathe have been avoided (\emph{e.g.} {\tt $\varphi$L3[1,2]} $\to \rut{CHl3[1,2]}$ or $\varphi \square \to \rut{CHbox}$).

\item Besides the two options to solve the RGEs avaliable in \dsixv{1.0} (exact numerical solution and leading logarithm), \dsixv{2.0} includes a third method, as the default setting. This method employs the Evolution Matrix approach, described in~\App{ap:evolution}. This method is numerically very precise and it is computationally faster than solving the RGEs exactly.

\item Many of the routines inherited from \dsixv{1.0} have changed names. For example, all routines related to the SMEFT now start with \rut{SMEFT...} and similarly for the LEFT (\emph{e.g.} \rut{SMEFTRunEGEs} and \rut{LEFTRunRGEs}), which makes it easier to use \mathe's autocompletion feature.
In addition, some routines in \dsixv{1.0} have been eliminated (or replaced by improved ones), and new routines have been implemented.
See~\Sec{subsec:sumroutines} for the complete list of routines in \dsixv{2.0}.

\item \dsixv{2.0} incorporates a reference repository of information about the SMEFT and the LEFT accessible through the routines \rut{SMEFTObjectList} and \rut{LEFTObjectList}, \rut{SMEFTOperators}  and \rut{LEFTOperators}, \rut{SMEFTParameterList} and \rut{LEFTParameterList}, \rut{ObjectInfo}, \rut{SMEFTOperatorsMenu}  and \rut{LEFTOperatorsMenu},  \rut{SMEFTOperatorsGrid} and \rut{LEFTOperatorsGrid}, and \rut{NIndependent}.
In addition, \dsixv{2.0} contains a full \mathe documentation system.

\item Setting the input values for the Wilson coefficients in the SMEFT or the LEFT through \rut{NewInput[\dots]}, \rut{ChangeInput[\dots]} or \rut{ReadInputFiles[\dots]} now checks the consistency of the given input, printing warnings when necessary. The same is done when setting scales through \rut{NewScale[...]}.
The input in \dsixv{2.0} is \emph{basis-independent}. See~\Sec{subsec:input} for details.
The user can also check the input values for the WCs at any time using the routines \rut{InputValues}, \rut{SMEFTLagrangian[HIGHSCALE]} or \rut{LEFTLagrangian[EWSCALE]}.

\item \dsixv{2.0} includes higher order corrections to matching coefficients and RG coefficients as compared to \dsixv{1.0}. In particular it includes SM beta functions up to five loops, and LEFT matching conditions in the SMEFT at one loop.

\end{itemize}


\section{Downloading, installing and loading \dsixbf}
\label{sec:down}

\dsix~is free software under the copyright of the \href{https://www.gnu.org/copyleft/gpl.html}{GNU General Public License}. There are two ways to download the package and install it:

\subsubsection*{Automatic installation}

The simplest way to download and install \dsix is to run the following command in a \mathe session:
\begin{lstlisting}[style=mathematica]
Import["https://raw.githubusercontent.com/DsixTools/DsixTools /master/install.m"];
\end{lstlisting}
\vspace{3mm}
This will download and install \dsix in the {\textit{Applications}} folder of the \mathe~base directory, activate the documentation and load the package. During the installation process, a pop up window will appear asking if you want to convert the .m files to .mx format. This option is recommended, since it significantly reduces the \dsix loading time.

\subsubsection*{Manual installation}

Alternatively, the user can also download and install \dsix manually. The package can be downloaded from the web page~\cite{dsixweb}:
\begin{center}
\href{https://dsixtools.github.io}{https://dsixtools.github.io}
\end{center}
We recommend placing the \dsix~folder inside the {\textit{Applications}} folder of \mathe's base directory, after which loading the package will be automatic.
Alternatively, the user can place the \dsix~folder in a different directory. In this case, loading the package will require specifying previously its location via
\begin{lstlisting}[style=mathematica]
pathtoDsixTools = "<directory>";
AppendTo[$Path, pathtoDsixTools];
\end{lstlisting}
\vspace{3mm}
As a final step, the user can activate the documentation by moving the contents of the zip file {\tt Documentation.zip} inside the {\tt DsixTools} folder, and applying
\begin{lstlisting}[style=mathematica]
If[$VersionNumber>=12.1,PacletDataRebuild[],RebuildPacletData[]];
\end{lstlisting}
inside a {\tt Mathematica} notebook.

\subsubsection*{Loading \dsix}

Once installed, the user can load \dsix~anytime with the command
\begin{lstlisting}[style=mathematica]
Needs["DsixTools`"]
\end{lstlisting}
\vspace{3mm}
When \dsix is loaded, a message is printed out with information about the version, the authors, and links to the relevant references and to the DsixTools website:\\[3mm] 
\framebox{
\includegraphics[width=16cm]{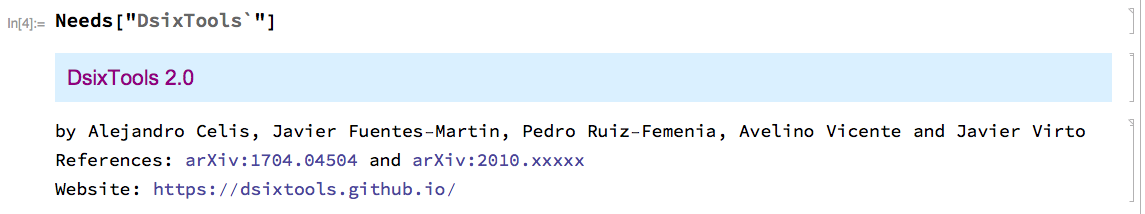}
}\\[4mm]
A typical loading time is about 5-10~s depending on the machine, if the .m to .mx conversion is done. When \dsix is loaded, several (relatively heavy) \mathe files containing SMEFT and LEFT beta functions, RGEs and evolution matrices, as well as the SMEFT-LEFT one-loop matching relations are loaded as well. This may be unnecessary for some \dsix applications. In this case the user can force \dsix to load without importing such files, by evaluating the line
\begin{lstlisting}[style=mathematica]
DsixTools`ImportFiles = False;
\end{lstlisting}
\vspace{3mm}
before loading \dsix.
This will reduce the loading time to under a second. If running or matching is required after loading \dsix in this mode, the corresponding files can be loaded by the user a posteriori, there is no need to reload \dsix.

\section{Using \dsixbf}
\label{sec:use}

In this Section we describe how to use \dsix in detail. After summarizing the \dsix routines and functions, the main features of the package will be explained with specific examples of use.

\subsection{A \dsixbf program}
\label{subsec:program}

The following is a simple but complete \dsix program which takes input from the user for the SMEFT Lagrangian at the UV scale $\Lambda_{\rm UV}=\rut{HIGHSCALE}$ and calculates the LEFT WCs at the IR scale $\Lambda_{\rm IR}=\rut{LOWSCALE}$, printing out one specific WC for illustration:
\vspace*{0.2cm}

\begin{lstlisting}[style=mathematica]
Needs["DsixTools'"]

NewScale[{HIGHSCALE -> 10000}];

NewInput[{Clq1[1,1,1,2] -> 1/HIGHSCALE^2, Clq1[1,1,2,1] -> 1/HIGHSCALE^2, CH -> -0.5/HIGHSCALE^2}];

RunDsixTools;

D6run[LeuVLL[2,2,1,1]] /. \[Mu] -> LOWSCALE
\end{lstlisting}

\vspace*{0.2cm}

\begin{figure}
  \centering
  \includegraphics[width=12cm]{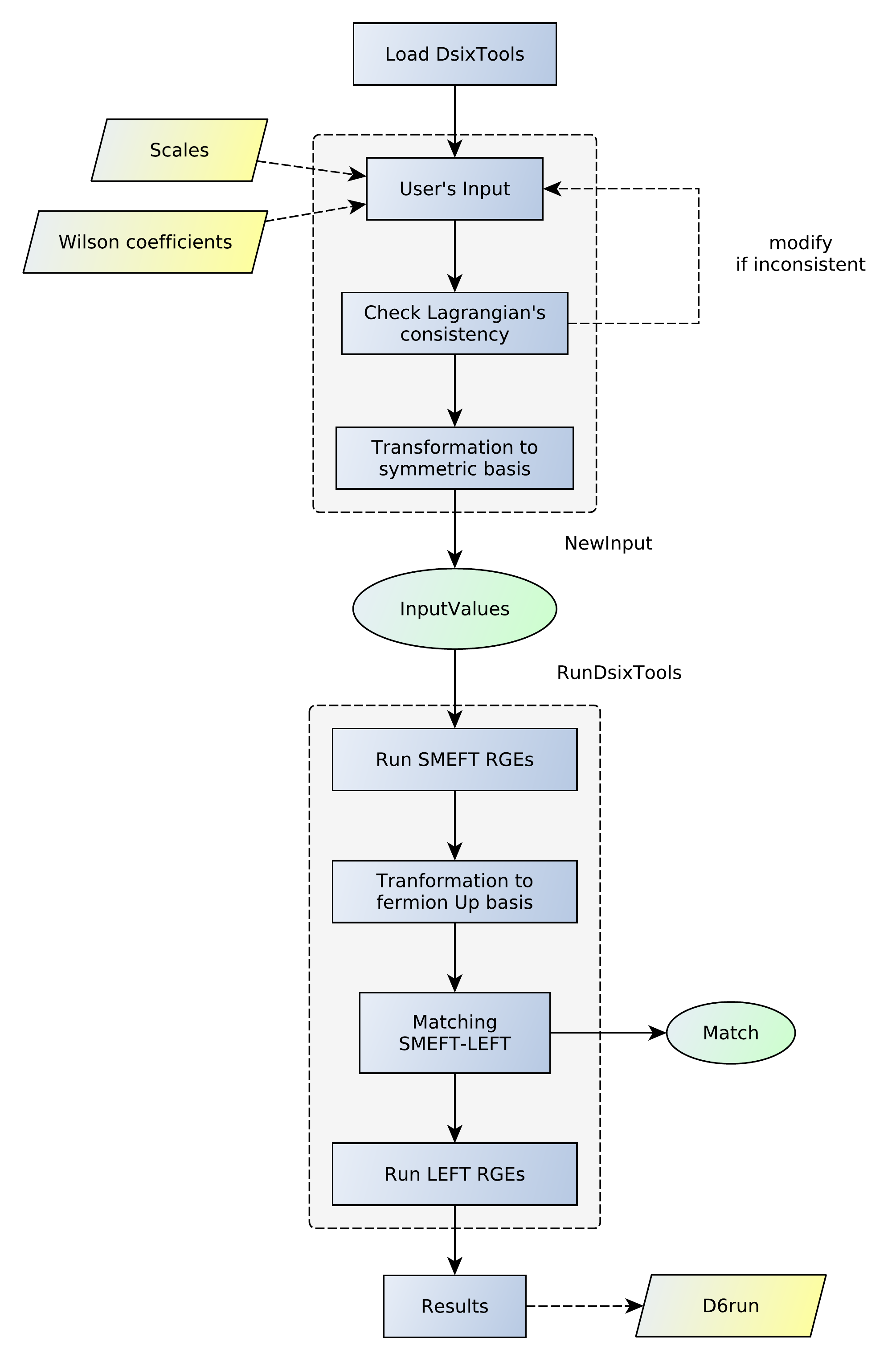}
  \caption{\small Example of a minimal \dsix program flowchart.}
  \label{fig:program}
\end{figure}

\noindent
The program begins by loading \dsix, as explained in Sec.~\ref{sec:down}. In the next line we provide the numerical value for the global variable \rut{HIGHSCALE} which corresponds to $\Lambda_{\rm UV}$ 
\begin{equation*}
\rut{HIGHSCALE} = \Lambda_{\rm UV} = 10 \, \text{TeV} \, . \\
\end{equation*}
In \dsix~{\bf all scales are given in GeV}. The third line defines the input values by means of the \rut{NewInput} \dsix routine. In this case the user is \emph{implicitly} specifying that the input WCs correspond to the SMEFT, and these take the values
\eq{
\left[ C_{\ell q}^{(1)} \right]_{1112} = \left[ C_{\ell q}^{(1)} \right]_{1121} = \, \, \frac{1}{\Lambda_{\rm UV}^2} \, \, = 10^{-8} \, \, \text{GeV}^{-2} \ ,\quad
C_\varphi = - \frac{0.5}{\Lambda_{\rm UV}^2} = - 5 \cdot 10^{-9} \, \, \text{GeV}^{-2} \, ,
}
at the new physics scale $\Lambda_{\rm UV} = 10$ TeV, with all the other WCs set to zero. We note that $[C_{\ell q}^{(1)}]_{1112} = [C_{\ell q}^{(1)}]_{1121}$ follows from the hermiticity of the Lagrangian, which implies the general relation $[C_{\ell q}^{(1)}]_{aabc} = [C_{\ell q}^{(1)}]_{aacb}^\ast$. If this condition were not respected by the arguments of the \rut{NewInput} routine, a message would be issued by \dsix and a modification of the input values in order to restore consistency would be applied (see~\Sec{subsec:input}). In the next line, the program makes use of the \rut{RunDsixTools} routine. This can be regarded as the \textit{master} \dsix routine, since it performs the three main tasks this package is designed for: it runs the SMEFT parameters from $\Lambda_{\rm UV}=\rut{HIGHSCALE}$ to $\Lambda_{\rm EW}=\rut{EWSCALE}$, matches to the LEFT, and finally runs the LEFT parameters from $\Lambda_{\rm EW}=\rut{EWSCALE}$ to $\Lambda_{\rm IR}=\rut{LOWSCALE}$.
The variable \rut{LOWSCALE} takes the default value \rut{LOWSCALE}$\,= 5 \, \text{GeV}$.
After evaluating \rut{RunDsixTools}, the \rut{D6run} function becomes available. The last line of the program precisely reads these results by printing the value of the LEFT WC $[ L_{eu}^{V,LL} ]_{2211}$ at $\mu = \Lambda_{\rm IR} = 5$ GeV, obtaining a numerical result
\begin{equation*}
\left[ L_{eu}^{V,LL} \right]_{2211} \simeq 6.22 \cdot 10^{-6} \, \, \text{GeV}^{-2} \, .
\end{equation*}

The general flowchart of this minimal program can be seen in Fig.~\ref{fig:program}. It clearly involves most of the main routines of \dsix and serves as an example of use in a practical scenario. However, some of the functionalities used in this program offer alternative possibilities and methods of application. For this reason, in the rest of the paper we explain in greater detail how to take full advantage of \dsix.

\subsection{Summary of \dsixbf routines}
\label{subsec:sumroutines}

Once the package has been loaded, the user can already execute all \dsix
functions and routines. Several \dsix global variables are also introduced
at this stage. Here we summarize the \dsix routines available to the user.\\

\subsubsection*{General variables and routines}

\begin{itemize}

\item \rut{DsixToolsVersion}: Returns the loaded version of \dsix.
\vspace{0.1cm}

\item \rut{DsixToolsDir}: Returns the directory holding the loaded version of \dsix.
\vspace{0.1cm}

\item \rut{HIGHSCALE}: UV scale  (in units of $\gev$) at which the SMEFT input is set and where the running in the SMEFT starts. 
\vspace{0.1cm}

\item \rut{EWSCALE}: Electroweak scale  (in units of $\gev$). This is the scale at which the LEFT input is set (either directly or through matching with the SMEFT), and where the running in the LEFT starts.
By default $\rut{EWSCALE}=91.1876\gev$, but it can be modified by means of \rut{NewScale} or \rut{NewInput}.
\vspace{0.1cm}

\item \rut{LOWSCALE}: IR scale (in units of $\gev$) which sets the lower limit beyond which the solution of the LEFT RGEs are only extrapolations. Since the LEFT in \dsixv{2.0} is the five-flavor theory, the default \dsix value is $\rut{LOWSCALE}=5$. 
\vspace{0.1cm}

\item \rut{RunDsixTools}: Master \dsix routine. It runs the SMEFT parameters from $\Lambda_{\rm UV}=\rut{HIGHSCALE}$ to $\Lambda_{\rm EW}=\rut{EWSCALE}$, matches to the LEFT, and then runs the LEFT parameters from $\Lambda_{\rm EW}$ to $\Lambda_{\rm IR}=\rut{LOWSCALE}$.
\end{itemize}

\subsubsection*{Reference}

\begin{itemize}

\item \rut{SMEFTObjectList} and \rut{LEFTObjectList}: List of SMEFT and LEFT {\it objects}, where an {\it object} is defined as a list of properties of a SMEFT or LEFT operator and its Wilson coefficients. 
\vspace{0.1cm}

\item \rut{SMEFTOperators} and \rut{LEFTOperators}: List of all SMEFT and LEFT operators in \dsix notation.
\vspace{0.1cm}

\item \rut{SMEFTParametersTotal} and \rut{LEFTParametersTotal}: List of all SMEFT and LEFT parameters (i.e. couplings, mass parameters and WCs) in \dsix notation.
\vspace{0.1cm}

\item \rut{SMEFTParameterList[<{\it attributes}>]} and \rut{LEFTParameterList[<{\it attributes}>]}: List of all {\it independent} SMEFT/LEFT parameters satisfying the condition given by {\it attributes}. The sequence of {\it attributes} can be chosen from the following predefined lists in the cases of the SMEFT and the LEFT, respectively:
\begin{align}
\text{SMEFT}\ : &\ {\tt \{``SM", ``D6", ``D5", ``D4", ``D2", ``BNV", ``BNC", ``LNV", ``LNC", ``CPodd",}
\nonumber\\
& {\tt ``CPeven", ``X3", ``H6", ``H4D2", ``X2H2", ``F2H3",``F2XH", ``F2H2D", ``LLLL",}
\nonumber\\
& {\tt ``RRRR", ``LLRR", ``LRLR", ``LRRL", ``B-violating", ``L-violating",}
\nonumber\\
& {\tt ``LFV", ``QFV", ``0F", ``2F", ``4F"\}}
\nonumber\\[2mm]
\text{LEFT}\ : &\ {\tt \{``QED\&QCD", ``D6", ``D5", ``D4", ``D3", ``D2", ``BNV", ``BNC", ``LNV", ``LNC",}
\nonumber\\
& {\tt  ``CPodd",``CPeven",``X3",``\nu\nu X",``LRX", ``LLLL", ``RRRR", ``LLRR", ``LRLR",}
\nonumber\\
& {\tt  ``LRRL", ``\Delta L=4", ``\Delta L=2", ``\Delta B=\Delta L=1", ``\Delta B=-\Delta L=1",}
\nonumber\\
& {\tt  ``LFV", ``QFV", ``0F", ``2F", ``4F"\}}
\nonumber
\end{align}
\rut{SMEFTParameterList[]} and \rut{LEFTParameterList[]} list all independent SMEFT and LEFT parameters.
\vspace{0.1cm}

\item \rut{SMEFTFindParameter[<{\it attributes}>,{\it parameter}]} and\\ \rut{LEFTFindParameter[<{\it attributes}>,{\it parameter}]}: Returns the position of {\it parameter} in the list \rut{SMEFTParameterList[{}]} or \rut{LEFTParameterList[{}]}. If the optional entry {\it attributes} is given, the position refers to the corresponding restricted list.
\vspace{0.1cm}

\item \rut{ObjectInfo[{\it parameter}]}: Prints information about {\it parameter}. \vspace{0.1cm}

\item \rut{SMEFTOperatorsMenu}, \rut{LEFTOperatorsMenu} and \rut{TotalOperatorsMenu}: Displays a clickable menu with information about the operators and parameters of the SMEFT and the LEFT.
\vspace{0.1cm}

\item \rut{SMEFTOperatorsGrid} and \rut{LEFTOperatorsGrid}: Creates a grid with all the SMEFT or LEFT operators. Moving the mouse on top of each entry displays the definition of the operator, and clicking on it a window with information about it is displayed.
\vspace{0.1cm}

\item \rut{SMEFTLagrangian[{\it scale}]} and \rut{LEFTLagrangian[{\it scale}]}:
Returns the SMEFT or LEFT Lagrangians at the renormalization scale given in the argument, corresponding to the current values given by \rut{InputValues}. If necessary, running and/or matching is performed internally.
\vspace{0.1cm}

\end{itemize}

\subsubsection*{Input \& Output}

\begin{itemize}

\item \rut{TurnOnMessages} and \rut{TurnOffMessages}: Turn on or off the messages written by the \dsix routines.
\vspace{0.1cm}

\item \rut{NewScale[\{{\it list}\}]}: Sets (or resets) the values of the scales indicated in {\it list}. For example, if
${\it list} = \{ \rut{HIGHSCALE} \to 5000 , \rut{LOWSCALE} \to 5 \}$
will set (or reset) $\Lambda_\text{UV}=5\tev$ and $\Lambda_\text{IR}=5\gev$.
\vspace{0.1cm}

\item \rut{InputValues}: Dispatch\footnote{
For those not familiar with \mathe dispatch tables, we clarify that these are lists (or tables) of pointers to replacements rules. In practice they work in exactly the same way as replacement rules, but their execution time is much lower when the list of replacements is long. It is possible to recover a normal replacement rule from the dispatch by applying to it the \mathe command {\tt Normal}.
}
that contains the current values of the Wilson coefficients at the input scale. It might refer to the SMEFT or the LEFT, depending on the last input defined by the user.

\item \rut{InputBasis}: Indicates the SMEFT flavor basis of the input in \rut{InputValues}. It can be {\tt``up''} or {\tt ``down''} (default).
\vspace{0.1cm}

\item \rut{NewInput[\{{\it list}\},<{\it additional}>]}: Resets the variable \rut{InputValues} putting to zero all $d>4$ WCs, and then replaces it by a new one in which the changes in {\it list} are applied. The optional \textit{additional} entries may also contain changes in the  scales \rut{HIGHSCALE}, \rut{EWSCALE} and \rut{LOWSCALE}, as well as in \rut{InputBasis}, e.g.,

\rut{NewInput[\{{\it list}\},HIGHSCALE->5000,InputBasis->"up"]}.
\vspace{0.1cm}

\item \rut{ChangeInput[\{{\it list}\}]}: Replaces (without resetting) the current dispatch \rut{InputValues} by a new one in which the changes in {\it list} are applied.
s\vspace{0.1cm}

\item \rut{SetSMLEFTInput}: Resets the variable \rut{InputValues} with the LEFT coefficients obtained from a matching to the SM.
\vspace{0.1cm}

\item \rut{ReadInputFiles[options\_file,\,\{WCsInput\_file\},\,\{SMInput\_file\},\{EFT\}]}: Reads all input files. \vspace{0.1cm}

\item \rut{WCXFtoSLHA[WCXF\_file,SLHA\_file,EFT]}: Translates the WCs file in WCxf format {\tt WCXF\_file} into an SLHA format file named {\tt SLHA\_file}. \vspace{0.1cm}

\item \rut{SLHAtoWCXF[SLHA\_file,WCXF\_file,SCALE,EFT]}: Translates the WCs file in SLHA format {\tt SLHA\_file} into an WCxf format file named {\tt WCXF\_file}.

\item \rut{AntisymmetryErrorsTotal}: List containing the accumulated set of errors fixed by \rut{NewInput}, \rut{ChangeInput} or \rut{ReadInputFiles} due to input not consistent with flavor-index symmetries.
\vspace{0.1cm}

\item \rut{NonHermitianErrorsTotal}: List containing the accumulated set of errors fixed by \rut{NewInput}, \rut{ChangeInput} or \rut{ReadInputFiles} due to input not consistent with hermiticity of the Lagrangian.
\vspace{0.1cm}

\end{itemize}

\subsubsection*{Operations with Wilson coefficients}

\begin{itemize}

\item \rut{D6Simplify[{\it expression}]}: Replaces  all  redundant  Wilson  coefficients  in  {\it expression}  by  their  expressions  in  terms  of  the  non-redundant ones. It also eliminates complex conjugates on real parameters.
\vspace{0.1cm}

\item \rut{SubRedundant}: Dispatch that replaces all redundant Wilson coefficients by their expressions in terms of the independent ones present in \rut{SMEFTParametersList[]} and \rut{LEFTParametersList[]}.
\vspace{0.1cm}

\item \rut{NIndependent[{\it parameter}]}: Returns the number of independent real parameters in {\it parameter}: 2 for a general complex parameter, 1 for a real parameter, and 0 for a redundant WC.
\vspace{0.1cm}

\item \rut{ToSymmetric[{\it X},{\it cat}]}: Returns {\it X} in the symmetric basis, where {\it X} is an object of category {\it cat} in array form.
\vspace{0.1cm}

\item \rut{ToSymmetricSingle[{\it parameter}]}: Returns the form of the SMEFT or LEFT {\it parameter} in the symmetric basis.
\vspace{0.1cm}

\item \rut{ToIndependent[{\it X},{\it cat}]}: Returns {\it X} in the independent basis, where {\it X} is an object of category {\it cat} in array form.
\vspace{0.1cm}

\item \rut{ToIndependentSingle[{\it parameter}]}: Returns the form of the SMEFT or LEFT {\it parameter} in the independent basis.
\vspace{0.1cm}

\item \rut{CheckAndSymmetrize[{\it X},{\it cat}]}: Returns {\it X} in the symmetric basis,  where {\it X} is an object of category {\it cat} in array form, after checking that all hermiticity and antisymmetry conditions are respected.  If any of the conditions are violated, some entries of {\it X} are modified.

\end{itemize}

\subsubsection*{SMEFT and LEFT running}

\begin{itemize}

\item \rut{RGEsMethod}: Indicates the method that \dsix is going to use to solve the RGEs. It is either $1$ (exact numerical solution), $2$ (first leading log) or $3$ (via the evolution matrix formalism). This variable is protected.
\vspace{0.1cm}

\item \rut{SetRGEsMethod[{\it n}]}: Sets the value of  \rut{RGEsMethod} to $n = 1,2$ or~$3$.
\vspace{0.1cm}

\item \rut{SMEFTLoopOrder}: Indicates the order that \dsix is going to use for the SM beta functions when running in the SMEFT.
The maximum (and default) in \dsixv{2.0} is $\rut{SMEFTLoopOrder}=5$. This variable is protected.
\vspace{0.1cm}

\item \rut{LEFTLoopOrder}: Indicates the order in QCD that \dsix is going to use for the strong coupling beta function and quark-mass anomalous dimensions when running in the LEFT.
The maximum (and default) in \dsixv{2.0} is $\rut{LEFTLoopOrder}=4$. This variable is protected.
\vspace{0.1cm}

\item \rut{SetSMEFTLoopOrder[{\it n}]} and \rut{SetLEFTLoopOrder[{\it n}]}: Set the values of \rut{SMEFTLoopOrder} and \rut{LEFTLoopOrder}
\vspace{0.1cm}

\item \rut{UseRGEsSM}: If $\rut{UseRGEsSM}=1$, \dsix will use the pure SM RGEs to run the SM parameters to the initial scale \rut{HIGHSCALE}.
\vspace{0.1cm}

\item \rut{$\beta$[{\it parameter}]}: Gives the beta function of the SMEFT or LEFT {\it parameter}.
\vspace{0.1cm}

\item \rut{$\beta$SM[{\it parameter}]}: Gives the SM beta function of the SM {\it parameter}.
\vspace{0.1cm}

\item \rut{SMEFTRunRGEs} and \rut{LEFTRunRGEs}: Solve the SMEFT and LEFT RGEs in each case.
\vspace{0.1cm}

\item \rut{D6run[{\it parameter},<"log10">]}: Gives the SMEFT or LEFT {\it parameter} as a function of the renormalization scale $\mu$. Including the optional argument {\tt "log10"} gives the function in terms of $t=\log_{10}(\mu/\gev)$.
\vspace{0.1cm}

\item \rut{SMEFTEvolve[{\it parameter},{\it final},{\it initial},<"log10">]}: Returns the SMEFT {\it parameter} at $\mu = {\it final}$ in terms of the SMEFT parameters at $\mu = {\it initial}$ using the evolution matrix method. 
\vspace{0.1cm}

\item \rut{LEFTEvolve[{\it parameter},{\it final},{\it initial},<"log10">]}:
Returns the LEFT {\it parameter} at $\mu = {\it final}$ in terms of the LEFT parameters at $\mu = {\it initial}$ using the evolution matrix method.
\vspace{0.1cm}

\item \rut{SMEFTrunnerExport[<{\it format}>,<{\it name}>]}: Exports the numerical values of the SMEFT parameters at the scale $\Lambda_{\rm EW}=\rut{EWSCALE}$ (after running). If no argument is given, \rut{SMEFTrunnerExport} generates a default output file named {\tt Output\_SMEFTrunner.dat}. This routine can also export the output to file with a {\it name} (without extension) and {\it format} chosen by the user (both arguments are required). The available formats are {\tt "SLHA"} (default \dsix format), {\tt "JSON"} and {\tt "YAML"}.
\vspace{0.1cm}

\item \rut{LEFTrunnerExport[<{\it format}>,<{\it name}>]}: Exports the numerical values of the LEFT parameters at the scale $\Lambda_{\rm IR}=\rut{LOWSCALE}$ (after running). If no argument is given, \rut{LEFTrunnerExport} generates a default output file named {\tt Output\_LEFTrunner.dat}. This routine can also export the output to file with a {\it name} (without extension) and {\it format} chosen by the user (both arguments are required). The available formats are {\tt "SLHA"} (default \dsix format), {\tt "JSON"} and {\tt "YAML"}. 
\vspace{0.1cm}

\end{itemize}

\subsubsection*{Matching at the EW scale}

\begin{itemize}

\item \rut{MatchingLoopOrder}: Indicates if the SMEFT-LEFT matching will be done at tree-level ($\rut{MatchingLoopOrder}=0$) or at one-loop ($\rut{MatchingLoopOrder}=1$). This variable is protected.
\vspace{0.1cm}

\item \rut{SetMatchingLoopOrder[{\it n}]}: Sets the value of \rut{MatchingLoopOrder} to $n$.
\vspace{0.1cm}

\item \rut{Match}: Dispatch that replaces all LEFT parameters by their numerical values at the matching scale, obtained after matching to the SMEFT.
\vspace{0.1cm}

\item \rut{MatchEW[{\it parameter}]}: Returns the matching condition of the LEFT {\it parameter} in terms of SMEFT parameters at the EW scale, in analytical form. 
\vspace{0.1cm}

\item \rut{MatchAnalytical}:
Dispatch that replaces all LEFT parameters by their analytical matching conditions, in terms of SMEFT parameters.
\vspace{0.1cm}

\item \rut{SMEFTLEFTMatch}:
Perfoms the matching between the SMEFT and the LEFT, at the order especified by \rut{MatchingLoopOrder}.
\vspace{0.1cm}

\item \rut{SMEFTRotateParameters[{\it scale}]}:
Returns a list containing two dispatches that transform the SMEFT parameters to the {\tt ``up''} and {\tt ``down''} bases at $\mu = scale$.
\vspace{0.1cm}

\item \rut{SMEFTToNewBasis[{\it basis},{\it scale}]}:
Dispatch that transforms the SMEFT parameters to a specific flavor basis ({\tt ``up''} or {\tt ``down''}) at $\mu=scale$.
\vspace{0.1cm}

\item \rut{LEFTToNewBasis[{\it scale}]}:
Dispatch that transforms the LEFT parameters to the mass basis at $\mu=scale$.
\vspace{0.1cm}

\item \rut{EWmatcherExport[<{\it format}>,<{\it name}>]}: Exports the numerical values of the LEFT parameters at the scale $\Lambda_{\rm EW}$ obtained after matching to the SMEFT. If no argument is given, \rut{EWmatcherExport} generates a default output file named {\tt Output\_EWmatcher.dat}. This routine can also export the output to file with a {\it name} (without extension) and {\it format} chosen by the user (both arguments are required). The available formats are {\tt "SLHA"} (default \dsix format), {\tt "JSON"} and {\tt "YAML"}. 
\vspace{0.1cm}

\end{itemize}

\subsubsection*{Other variables and routines}

\begin{itemize}

\item \rut{Biunitary[{\it matrix}]}: Applies a biunitary transformation diagonalizing the square {\it matrix}, and provides the rotation matrices and the eigenvalues.
\vspace{0.1cm}

\item \rut{LoopParameter}: Appears in analytical expressions such as beta functions and matching conditions, separating different loop orders. An $n$-loop term is proportional to $(\rut{LoopParameter})^n$ (except for $n=1$ in the beta functions). This variable is protected.

\vspace{0.1cm}

\end{itemize}

\subsection{Input values in \dsix}
\label{subsec:input}

One of the first steps in every \dsix program is to define the input. This includes the numerical values of the SMEFT or LEFT parameters at the input scale, the relevant scales for matching and RGE running ($\Lambda_{\rm UV}=\rut{HIGHSCALE}$, $\Lambda_{\rm EW}=\rut{EWSCALE}$ and $\Lambda_{\rm IR}=\rut{LOWSCALE}$), and some \dsix options. The input values for the SM parameters, which are used by default and in the evolution matrix method, are given in Table~\ref{tab:inputs}. 

\begin{table}[t]
\renewcommand{\arraystretch}{1.4}
\caption{Default \dsix inputs for the SM parameters at the scale $M_Z = 91.1876\gev$.}\label{tab:inputs}
\begin{center}
\begin{tabular}{|cc|}
\hline
Parameter & Value\\
\hline
$g$ & $0.6515$\\
$g^\prime$ & $0.3576$\\
$g_s$ & $1.220$\\
$\lambda$ & $0.2813$\\
$m^2$ & $8528~\mathrm{GeV}^2$\\
$\Gamma_u$  & $\begin{pmatrix} 7.109\times10^{-6} & -8.175\times10^{-4} & (8.176 + 3.265\, i)\times 10^{-3}\\ 1.636\times10^{-6} & 3.551\times10^{-3} & -4.017\times10^{-2} \\ 
(0.782 + 2.522\, i)\times10^{-8} & 1.540\times10^{-4} & 0.970\end{pmatrix}$\\
$\Gamma_d$ & $\mathrm{diag}(1.551\times10^{-5},3.165\times10^{-4},1.637\times10^{-2})$\\
$\Gamma_e$ & $\mathrm{diag}(2.944\times10^{-6},6.071\times10^{-4},1.021\times10^{-2})$\\
$\theta_s,\theta,\theta^\prime$ & 0\\
\hline
\end{tabular}
\end{center}
\end{table}

There are two ways of defining an input. The first way, which we call {\bf notebook input}, is to introduce the input values directly in the \mathe notebook. This is the method used in the example program shown in \Sec{subsec:program}. Alternatively, the user can also set the input by reading external files containing the input values. We will refer to this approach as {\bf external files input}. We now explain these two approaches and how to use them. For definiteness, we will concentrate on the SMEFT. For setting input in the LEFT, the steps and routines are completely analogous.

\subsubsection*{Notebook input}

The simplest way of setting the input in \dsix is to introduce the values directly in the \mathe notebook. The \dsix options and the relevant scales for the RGE running can be introduced easily. For instance,
\begin{lstlisting}[style=mathematica]
UseRGEsSM = 0;
NewScale[{HIGHSCALE->10000}];
\end{lstlisting}
would set the \rut{UseRGEsSM} option to 0 and the high-energy scale $\Lambda_{\rm UV}=10\tev$. The SMEFT or LEFT parameters (including the SM or QCD \& QED inputs) can be introduced by means of the \rut{NewInput} routine. This routine resets the input so that the WCs take their default values and then applies the changes indicated by the user.\footnote{The default SMEFT and LEFT values correspond to the SM and QED\&QCD benchmarks, respectively, in both cases with all Wilson coefficients of Dimension-five and -six operators set to zero and default values for the coefficients of Dimension $\leq 4$ operators.} For instance, the program of Sec.~\ref{subsec:program} includes the line
\begin{lstlisting}[style=mathematica]
NewInput[{Clq1[1,1,1,2] -> 1/HIGHSCALE^2, Clq1[1,1,2,1] -> 1/HIGHSCALE^2, CH -> -0.5/HIGHSCALE^2}];
\end{lstlisting}
which, as discussed already, sets $[ C_{\ell q}^{(1)} ]_{1112} = [ C_{\ell q}^{(1)} ]_{1121} = 1 / \Lambda_{\rm UV}^2 = 10^{-8}$ GeV$^{-2}$ and $C_\varphi = - 0.5 / \Lambda_{\rm UV}^2 = - 5 \cdot 10^{-9}$ GeV$^{-2}$, if the new physics scale $\Lambda_{\rm UV}$ is previously set to $10$ TeV. We note that only the non-vanishing WCs must be given and the rest are assumed to be zero.

As explained in Appendix~\ref{ap:parameters}, some of the 2- and 4-fermion operators in the SMEFT and the LEFT possess specific symmetries under the exchange of flavor indices. 
In particular, these symmetries imply conditions to be enforced in the input WCs in order to avoid two types of inconsistencies:
\begin{enumerate}

\item {\bf Hermiticity:} The hermiticity of the Lagrangian imposes certain conditions on some WCs, and these must be respected by the input provided by the user. For instance, an input with $[ C_{\ell q}^{(1)}]_{1112} \ne [ C_{\ell q}^{(1)}]_{1121}^\ast$ would be inconsistent.

\item {\bf Antisymmetry:} Some LEFT operators are antisymmetric under the exchange of two flavor indices and thus vanish. For practical reasons, we have not excluded these operators from the WC input list, but rather require that the corresponding WCs vanishes. For instance, an input with $[ L_{\nu \gamma}]_{11} \ne 0$ would be inconsistent.

\end{enumerate}

In order to avoid potential issues associated to inconsistent inputs, \dsix includes user-friendly input routines that simplify the user's task. \dsix accepts input values for the WCs of any set of operators (belonging to the Warsaw or San Diego bases) and then
checks for possible consistency problems. When the user's input is not consistent, a warning is issued and \dsix corrects the input by replacing it by a new one that ensures a complete consistency of the Lagrangian. For instance, this would be case if the user initializes \rut{HIGHSCALE} and then runs
\begin{lstlisting}[style=mathematica]
NewInput[{Clq1[1,1,1,2] -> 1/HIGHSCALE^2}];
\end{lstlisting}
since this command sets $[ C_{\ell q}^{(1)}]_{1112} = 1 / \Lambda_{\rm UV}^2$ and $[ C_{\ell q}^{(1)}]_{1121} = 0 \ne [ C_{\ell q}^{(1)}]_{1112}^\ast$. The list of invalid input values can be seen by clicking on a button named {\tt Input errors} that appears after running \rut{NewInput}. \dsix fixes this inconsistency by defining $\mathcal{L} = \frac{1}{2} \left(\mathcal{L}_{\rm in} + \mathcal{L}_{\rm in}^\ast\right)$, where $\mathcal{L}_{\rm in}$ is the input Lagrangian containing the inconsistency.\,\footnote{
Even though this correction is only applied when the input Lagrangian is not Hermitian, we note that in case of a consistent input this change would have no effect.} 
The resulting input values after this correction are $[ C_{\ell q}^{(1)}]_{1112} = [ C_{\ell q}^{(1)}]_{1121} = 1 / ( 2 \Lambda_{\rm UV}^2 )$, now satisfying $[ C_{\ell q}^{(1)}]_{1121} = [ C_{\ell q}^{(1)}]_{1112}^\ast$.
The hermiticity correction only needs to be applied to those operators for which we do not need to add explicitly its hermitian conjugates in the Lagrangian because they are already included among their flavor components.
We finally note that in some cases other Wilson coefficients, related to these by the two reasons given above, are also modified. The user should therefore pay attention to these messages to make sure that the input has been correctly introduced.

Furthermore, \dsix transforms all WCs to the \textit{symmetric basis}, defined as the basis in which the WCs follow the same symmetry conditions as the associated operators. We refer to Appendix~\ref{ap:bases} for more information about this basis. For example, in the symmetric basis $[C_{\ell\ell}]_{1122} = [C_{\ell \ell}]_{2211}$ since $[Q_{\ell\ell}]_{1122} = [Q_{\ell \ell}]_{2211}$. This is the basis used internally by \dsix. Nevertheless, the user needs not to worry about this, {\bf since the input is always unambiguous}.
In fact, this is one of the virtues of the input system in \dsixv{2.0}: the user introduces directly a \emph{Lagrangian}, which as such is basis-independent, e.g.,
\begin{lstlisting}[style=mathematica]
NewInput[{Cll[1,1,2,2] -> x, Cll[2,2,1,1] -> y}];
\end{lstlisting}
sets the input SMEFT Lagrangian
\eq{
{\mathcal L}_{\rm SMEFT} =  {\mathcal L}_{\rm SM}
+ x\,[Q_{\ell\ell}]_{1122} + y\,[Q_{\ell\ell}]_{2211}\ ,
}
which is unambiguous, and understood by \dsix with no regard to the index symmetry relation $[Q_{\ell\ell}]_{1122} = [Q_{\ell \ell}]_{2211}$.

After defining the input values with the \rut{NewInput} routine the dispatch \rut{InputValues} gets (re)initialized. This dispatch can be used to print the input value of any SMEFT or LEFT parameter. For instance, after running
\begin{lstlisting}[style=mathematica]
NewInput[{Cll[1,1,2,2] -> 10^(-8)}];
\end{lstlisting}
one can evaluate
\begin{lstlisting}[style=mathematica]
Cll[1,1,2,2] /. InputValues
\end{lstlisting}
and obtain the result $5 \cdot 10^{-9}$ GeV$^{-2}$. This is the input value given with the \rut{NewInput} routine to the SMEFT WC $[ C_{\ell \ell} ]_{1122}$, after transforming to the symmetric basis. In this basis $[ C_{\ell \ell}]_{2211} = [ C_{\ell \ell} ]_{1122}$, and due to $[ Q_{\ell \ell} ]_{2211} = [ Q_{\ell \ell}]_{1122}$ this is equivalent to the input given by the user:
\begin{align*}
\text{User's input:} \quad & \left[ C_{\ell \ell} \right]_{1122} = 10^{-8} \, \, \text{GeV}^{-2} \quad \text{and} \quad \left[ C_{\ell \ell} \right]_{2211} = 0 \, , \\
\text{In symmetric basis:} \quad & \left[ C_{\ell \ell} \right]_{1122} = 5 \cdot 10^{-9} \, \, \text{GeV}^{-2} \quad \text{and} \quad \left[ C_{\ell \ell} \right]_{2211} = 5 \cdot 10^{-9} \, \, \text{GeV}^{-2} \, .
\end{align*}
This can be clearly seen by evaluating the command
\begin{lstlisting}[style=mathematica]
MCll /. InputValues
\end{lstlisting}
which prints the complete $C_{\ell \ell}$ WC in array form.
The input values in the \textit{independent} basis (see~\App{ap:bases})
can be obtained by applying the routine \rut{ToIndependent}:
\begin{lstlisting}[style=mathematica]
ToIndependent[MCll,6] /. InputValues
\end{lstlisting}
which in this case results in the same input introduced before since $[ C_{\ell \ell} ]_{1122}$ is one of the independent WCs.

Finally, once the input values have been set, the user can change them individually at any moment in the notebook. This is done with the \rut{ChangeInput} routine. In contrast to \rut{NewInput}, this routine does not reset the input to default values, but just applies the changes demanded by the user. For instance,
\begin{lstlisting}[style=mathematica]
ChangeInput[{CHG -> 10^(-6)}]
\end{lstlisting}
would change the value of $C_{\varphi G}$ to $10^{-6}$ GeV$^{-2}$ in the current \rut{InputValues} dispatch, without altering the values of the other SMEFT parameters. Exactly as \rut{NewInput}, the \rut{ChangeInput} routine also checks the consistency of the input Lagrangian provided by the user and then translates the 2- and 4-fermion WCs to the symmetric basis.

\subsubsection*{External files input}

Alternatively, the user can set the program options and provide input values from external files. This is done with the \rut{ReadInputFiles} routine. For instance,
\begin{lstlisting}[style=mathematica]
ReadInputFiles["Options.dat","WCsInput.dat","SMInput.dat", "SMEFT"]}
\end{lstlisting}
applies the content of three SMEFT input files.\footnote{The use of input files for the LEFT is completely analogous, the only difference being that instead of input values for the SM parameters one must provide input values for the QCD \& QED parameters, and that the last option should be "LEFT" instead of "SMEFT".} The file {\tt Options.dat} contains the option values to be used in the program, the file {\tt WCsInput.dat} contains the input values for the SMEFT WCs at $\mu=\Lambda_{\rm UV}$, and the file {\tt SMInput.dat} contains the input values for the SM parameters. Examples for all of these files (and the corresponding ones for the LEFT) can be found in the {\tt IO} folder of \dsix. Each of the entries in these files are accompanied by comments that make them self-explanatory. Similarly to the case of notebook input, the \rut{InputValues} dispatch gets initialized and can be used after using \rut{ReadInputFiles}.

The default \dsix input and output format~is inspired by the Supersymmetry Les Houches Accord (SLHA)~\cite{Skands:2003cj,Allanach:2008qq}. Input files are distributed in blocks, each devoted to a set of parameters. Any complex parameter is given in two blocks, so that real and imaginary parts should be provided separately. Furthermore, WCs carrying flavor indices should be provided individually for each flavor combination. Analogously to the notebook input case, all WCs are assumed to vanish by default. Therefore, it suffices to include the non-zero WCs (and only these) in the input card.  Furthermore, the routine \rut{ReadInputFiles} will also check that the set of input values provided by the user is consistent. If any of the hermiticity or antisymmetry conditions on the WCs are not satisfied, a message will be issued and the corresponding input values modified in order to restore consistency.

Additionally, \dsix can also read WCs input files in WCxf format~\cite{Aebischer:2017ugx}, a standard data exchange format for numerical values of Wilson coefficients. In this case, the WCs input card can be a \json or \yaml file. Note however that reading \yaml input files requires previous installation of a \yaml importer for \mathe~\cite{myaml}. For more details about the WCxf format, such as the specific fermion basis that is implicitly assumed, we refer to~\cite{Aebischer:2017ugx}.

\subsection{RGE running}
\label{subsec:rge}

Once the initial conditions at some energy scale $\Lambda_{\rm start}$ are defined, the user can apply the RGEs to obtain the resulting Lagrangian parameters at the different energy scale $\Lambda_{\rm end}$. The SMEFT running between $\Lambda_{\rm UV}=\rut{HIGHSCALE}$ and $\Lambda_{\rm EW}=\rut{EWSCALE}$ is performed with the \rut{SMEFTRunRGEs} routine, while the LEFT running between $\Lambda_{\rm EW}=\rut{EWSCALE}$ and $\Lambda_{\rm IR}=\rut{LOWSCALE}$ is performed with the \rut{LEFTRunRGEs} routine.
Alternatively, the user can also perform the full RGE evolution from $\Lambda_{\rm UV}>\Lambda_{\rm EW}$ down to $\Lambda_{\rm IR}<\Lambda_{\rm EW}$ by means of the \rut{RunDsixTools} master routine, which internally makes use of \rut{SMEFTRunRGEs} and \rut{LEFTRunRGEs} and also applies the SMEFT-LEFT matching at $\Lambda_{\rm EW}$ with \rut{SMEFTLEFTMatch}. 

\dsix has three different methods for the resolution of the RGEs, which the user can choose by setting the flag \rut{RGEsMethod}:

\begin{itemize}

\item \emph{``Exact''} ($\rut{RGEsMethod}=1$): This method applies the \mathe internal command {\tt NDSolve} for the numerical resolution of the differential equations. Given the large number of differential equations involved in this case (several thousands), this might be time consuming, with each evaluation requiring a few (< 10) seconds, the exact number depending on the particular case and computer.

\item \emph{``First leading log''} ($\rut{RGEsMethod}=2$): This approximate method might be sufficient for many phenomenological studies, in particular when the initial and final scales are not too far from each other. The solution of the RGEs is obtained as
\begin{equation}
C_i(\mu) = C_i(\Lambda_{\rm start}) + \frac{\beta_i}{16 \pi^2} \log \left( \frac{\mu}{\Lambda_{\rm start}} \right)\,,
\end{equation}
where $C_i$ is any of the running parameters, $\mu$ is the renormalization scale and $\beta_i$ is the beta function for the $C_i$ parameter evaluated at $\mu = \Lambda_{\rm start}$. This method is much faster but neglects leading log resummation.

\item \emph{``Evolution matrix''} ($\rut{RGEsMethod}=3$): This method uses an evolution matrix formalism, explained in detail in Appendix~\ref{ap:evolution}.

\end{itemize}

By default, the SM parameters are assumed to be given at the electroweak scale $\Lambda_{\rm EW}=M_Z=91.1876\gev$. Therefore, before running down from $\Lambda_{\rm UV}$ to $\Lambda_{\rm EW}$ they must be computed at $\Lambda_{\rm UV}$. In case the user chooses to solve the RGEs with \rut{RGEsMethod}=1 (NDSolve) or \rut{RGEsMethod}=2 (leading log), this can be done by running up from the electroweak scale using pure SM RGEs, hence neglecting possible deviations caused by non-zero SMEFT WCs.\footnote{The user can check the validity of this approximation by using the \dsix routines, for instance by checking whether the resulting values for the SM parameters at the electroweak scale (after running down) do not match their initial values. This can be fixed by readjusting the SM parameters at $\Lambda_{\rm UV}$. We note, however, that one should also take into account NP corrections to the standard electroweak parameters induced by non-zero SMEFT WCs.} However, in case the user prefers to give the SM parameters directly at the high-energy scale $\Lambda_{\rm UV}$, this can be done by setting the \rut{UseRGEsSM} option to $0$. This choice is recommended when the user wants to use the \emph{First leading log} method to solve the RGEs. In the case the user chooses \rut{RGEsMethod}=3 (\dsix default) for the resolution of the RGEs (the evolution matrix method), this is implicitly taken into account. Our derivation of the evolution matrix already enforces the SM parameters to be fixed to their measured values at the EW scale.

The user chooses between these three methods by setting the global option \rut{RGEsMethod} to 1 (for the \emph{``Exact''} method), to 2 (for the \emph{``First leading log''} method) or to 3 (for the \emph{``Evolution matrix''} method), via the routine \rut{SetRGEsMethod}. After running, the results are saved in the function~\rut{D6run}, such that \rut{D6run[{\it parameter}]} returns the parameter {\it parameter} after RGE running as a function of the renormalization scale $\mu$. Therefore, the user can easily read the results by running commands such as
\begin{lstlisting}[style=mathematica]
D6run[Clq1[2,2,3,3]] /. \[Mu] -> EWSCALE
\end{lstlisting}
\vspace{1mm}
which would give the result for $[C_{\ell q}^{(1)}(\Lambda_\text{EW})]_{2233}$.

The results obtained after running can be also exported to a text file. This is done with the routines \rut{SMEFTrunnerExport[]} and \rut{LEFTrunnerExport[]}, which generates the files {\tt Output\_SMEFTrunner.dat} or {\tt Output\_LEFTrunner.dat} in each case, with SLHA format (completely analogous to the WCs input card in this format).
Alternatively, the user can export the results into text files following the WCxf convention~\cite{Aebischer:2017ugx} by adding an argument to the previous routines: \rut{SMEFTrunnerExport[{\it format}]} and \rut{LEFTrunnerExportWCXF[{\it format}]}, with {\it format} being {\tt JSON} or {\tt YAML}.

\bigskip

The evolution matrix method is also used internally by default when evaluating the routines \rut{SMEFTEvolve} and \rut{LEFTEvolve}. 
These routines provide a semi-analytical solution of the RGEs. For example,
\begin{lstlisting}[style=mathematica]
EvolveSMEFT[CdG[2,2], EWSCALE, HIGHSCALE]
\end{lstlisting}
\vspace{1mm}
returns an analytical expression for the SMEFT WC $\left[C_{dG} \right]_{22}$ at $\mu = \Lambda_{\rm EW}$ as a function of the SMEFT parameters at $\mu = \Lambda_{\rm UV}$, with numerical coefficients.
This easily allows the user to identify the most relevant contributions to the running, as well as running fast numerical scans of the EFT parameter space.
    
\bigskip    
    
Finally, we point out that \dsix can also be used for analytical calculations involving the SMEFT or LEFT beta functions, since these are available to the user right after loading the package. They can be printed simply by evaluating \rut{$\beta$[parameter]}, where {\tt parameter} must be a valid SMEFT or LEFT parameter (a member of \rut{SMEFTParameterListTotal} or \rut{LEFTParameterListTotal}).
For instance, \rut{$\beta$[LdddSRR[2, 3, 3, 3]]} returns the beta function of the LEFT WC $[ L_{ddd}^{S,RR}]_{2333}$.

\subsection{SMEFT-LEFT matching at the electroweak scale}
\label{subsec:matching}

In the first step of the matching process, \dsix transforms all the SMEFT parameters at the EW scale to the \textit{up basis}, applying the required biunitary transformations to the fermion mass matrices (which include contributions from dimension-six operators).
The up basis, defined in~\App{sec:SMEFT}, allows one to properly identify the top quark, one of the fields that decouples in the matching. After this transformation, the LEFT parameters at the electroweak scale are computed, using either the full tree-level matching~\cite{Jenkins:2017jig}\,\footnote{
We have independently derived the tree-level results in two different ways, finding full agreement.
} (if $\rut{MatchingLoopOrder}=0$) or the full one-loop matching~\cite{Dekens:2019ept} (if $\rut{MatchingLoopOrder}=1$). In order to set the value of \rut{MatchingLoopOrder} prior to the matching procedure, the user can use the routine \rut{SetMatchingLoopOrder}.
The result of the matching of the LEFT coefficients at the EW scale is given in the tree-level mass basis.

The SMEFT-LEFT matching is performed by evaluating
\begin{lstlisting}[style=mathematica]
SMEFTLEFTMatch;
\end{lstlisting}
\vspace{1mm}
This routine (re)initializes the \rut{Match} dispatch, which can be used to obtain the numerical values of the LEFT parameters after the matching at the electroweak scale. Therefore
\begin{lstlisting}[style=mathematica]
LeeVLL[1,1,1,1] /. Match
\end{lstlisting}
\vspace{1mm}
would return the numerical value of $[ L_{ee}^{V,LL}(\Lambda_\text{EW})]_{1111}$ in units of $\gev^{-2}$. The corresponding analytical expressions can be obtained by using \rut{MatchEW}, e.g.
\begin{lstlisting}[style=mathematica]
LeeVLL[1,1,1,1] // MatchEW
\end{lstlisting}
\vspace{1mm}
Note that \rut{MatchEW} does not require running \rut{SMEFTLEFTMatch}.

Since the LEFT is more general than the SMEFT low-energy limit, not all the LEFT operators are generated from a matching to the SMEFT. For instance, applying the command
\mbox{\tt ML$\nu \gamma$ /.\rut{MatchAnalytical}} would return a $3 \times 3$ matrix full of zeros, since the LEFT operator $\Op_{\nu \gamma}$ is not present in the SMEFT. 

Furthermore, as explained, the first step of the routine \rut{MatchSMEFTLEFT} is to rotate all SMEFT parameters to the fermion up basis. These rotations can be readily obtained by means of the \rut{SMEFTRotateParameters} routine by evaluating e.g.,
\begin{lstlisting}[style=mathematica]
{ToUpBasis, ToDownBasis} = SMEFTRotateParameters[EWSCALE];
\end{lstlisting}
This will create the dispatches {\tt ToUpBasis} and {\tt ToDownBasis}, which can be used to obtain any SMEFT parameter in the up and down bases at the electroweak scale. For instance,
\begin{lstlisting}[style=mathematica]
CuH[1,2] /. ToUpBasis
CuH[1,2] /. ToDownBasis
\end{lstlisting}
would return $[ C_{u \varphi} ]_{12}$ in GeV in the up and down bases at $\Lambda_{\rm EW}$. We also note that the \rut{SMEFTRotateParameters} routine can be used to obtain the SMEFT parameters in the up and down bases at any scale $\mu \geq \Lambda_\text{EW}$. For instance, running
\begin{lstlisting}[style=mathematica]
{ToUpBasis, ToDownBasis} = SMEFTRotateParameters[500];
\end{lstlisting}
creates the dispatches {\tt ToUpBasis} and {\tt ToDownBasis}, now applicable to obtain any SMEFT parameter in the up and down bases at $500$ GeV. Finally, if the user is interested in only one of the two fermion bases, up or down, the routine to be used is \rut{SMEFTToNewBasis}. For instance,
\begin{lstlisting}[style=mathematica]
ToUpBasis = SMEFTToNewBasis["up",EWSCALE];
\end{lstlisting}
would only create the {\tt ToUpBasis} dispatch.

\bigskip

All these results can be exported to external text files with the routine \rut{EWmatcherExport}. 
This generates the file {\tt Output\_EWmatcher.dat}, in SLHA format.
The results can also be exported in WCxf convention by adding two arguments ({\it format} and {\it name}) to the previous routine: \rut{EWmatcherExport[{\it format},{\it name}]}, with {\it format} being {\tt "JSON"} or {\tt "YAML"}.
The resulting file will always be in the up basis, denoted as {\tt Warsaw Up} basis in the WCxf exchange format documentation~\cite{Aebischer:2017ugx}.

\subsection{Reference guide and tools in \dsix}
\label{subsec:dictionary}

\dsix aims at a simple and visual experience. This is accomplished via a variety of routines, some of which grant the user simple access to the most basic, useful and comprehensive information about the LEFT and the SMEFT, while others implement practical operations on the Wilson coefficients.

The first repository of information available is contained in the variables \rut{SMEFTObjectList} and \rut{LEFTObjectList}, which are lists of certain \emph{objects}, one for each operator of the EFT (75 for the SMEFT, 103 for the LEFT, up to dimension six). Each object is itself a list containing: the flavor matrix of WCs, the name of the (head of) the WCs, the name of the operator, the symmetry category, the flavor dimension, the canonical dimension, the EFT, the operator class, the broken symmetry (if any), and the \LaTeX\ form for both the operator and its definition. A flattened list of all the parameters appearing in the first position of the objects in \rut{SMEFTObjectList} and \rut{LEFTObjectList} is given in \rut{SMEFTParametersTotal} and \rut{LEFTParametersTotal}:
\begin{align}
{\tt SMEFTParametersTotal} &= {\tt Flatten[SMEFTObjectList[[All,1]]]} 
\nonumber\\[2mm]
{\tt LEFTParametersTotal} &= {\tt Flatten[LEFTObjectList[[All,1]]]} 
\nonumber
\end{align}
which are all the parameters that might receive input values or output results.
However, not all these parameters are independent, and not all are complex-valued. The function \rut{NIndependnet[{\it parameter}]} returns the number of independent real parameters in a given parameter: 2 for a general complex parameter, 1 for a real parameter and 0 for a redundant one (as chosen by \dsix convention). The list of independent parameters are contained in the lists \rut{SMEFTParameterList[]} and \rut{LEFTParameterList[]}, which match the operators in \rut{SMEFTOperators} and \rut{LEFTOperators}. For example the LEFT Lagrangian in the ``independent basis'' (containing only non-redundant operators) is given by
\begin{lstlisting}[style=mathematica]
LEFTParameterList[].LEFTOperators
\end{lstlisting}
In addition, in order to find the position that a \emph{parameter} occupies in \rut{SMEFTParameterList[]} or \rut{LEFTParameterList[]} one can use the routines \rut{SMEFTFindParameter[{\it parameter}]} and \rut{LEFTFindParameter[{\it parameter}]}.

The routines \rut{SMEFTParameterList} and \rut{LEFTParameterList} also admit arguments in order to choose subsets of parameters with certain properties. For example
\begin{lstlisting}[style=mathematica]
SMEFTParameterList["D6","LNV"]
\end{lstlisting}
lists the non-redundant Lepton-Number-violating SMEFT parameters of canonical dimension six. For a list of attributes that can be chosen as arguments in \rut{SMEFTParameterList} and \rut{LEFTParameterList} see~\Sec{subsec:sumroutines}.

\begin{figure}
  \centering
  \includegraphics[width=6.5cm]{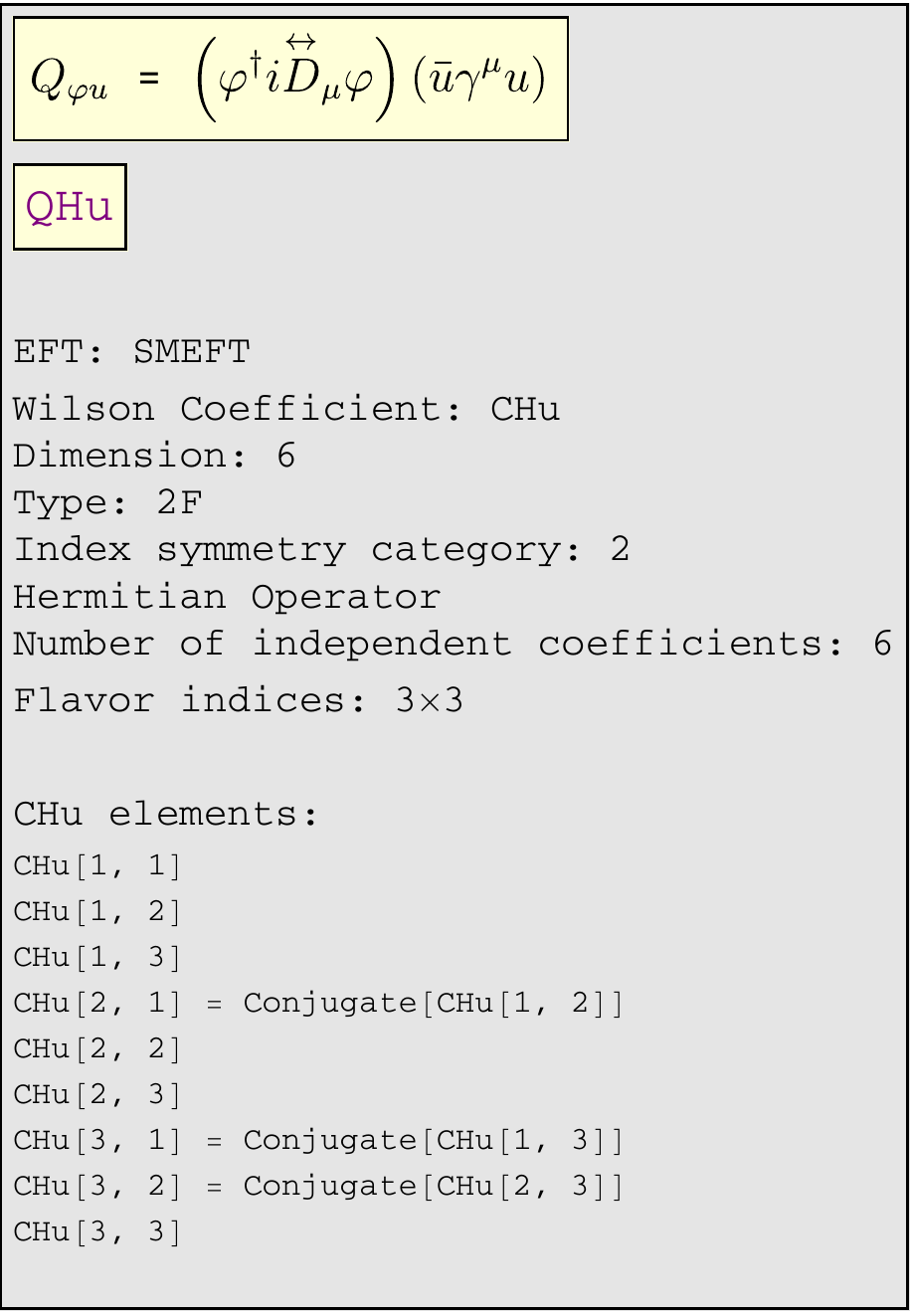}
  \caption{\small Information about the SMEFT WC $C_{\varphi u}$ obtained after evaluating \rut{ObjectInfo[CHu]}, or by using the interfaces \rut{SMEFTOperatorsMenu} or \rut{SMEFTOperatorsGrid}.}
  \label{fig:dict}
\end{figure}

\begin{figure}
  \centering
  \includegraphics[width=16.5cm,height=5.1cm]{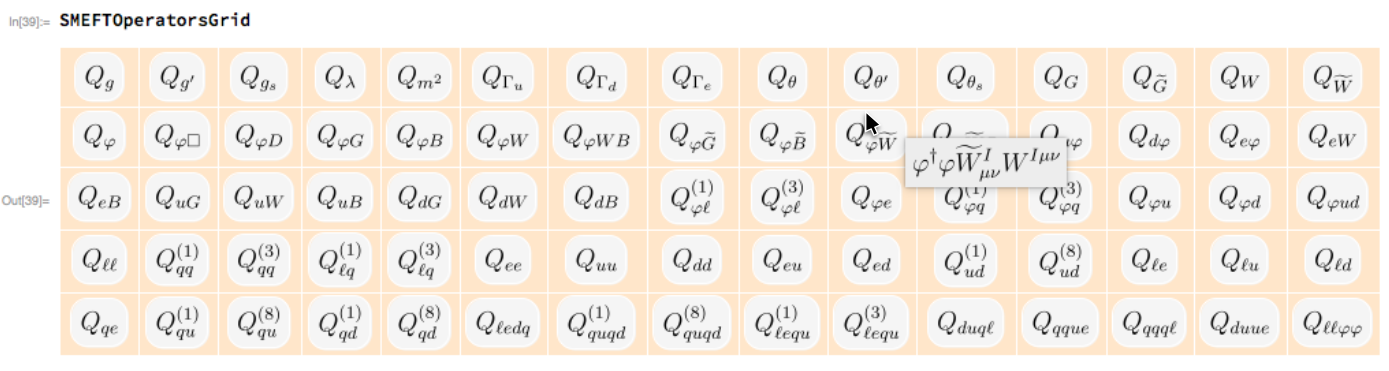}
  \caption{\small Result of evaluating \rut{SMEFTOperatorsGrid}. 
  Positioning the mouse on top of any operator displays its definition, and clicking on it opens a pop-up containing the corresponding chart of Fig.~\ref{fig:dict}.
  This grid can be saved in a notebook and used later in a fresh Kernel without loading \dsix.}
  \label{fig:grid}
\end{figure}

More visual information on the properties of operators and parameters can be obtained via a series of new routines. The routine \rut{ObjectInfo} displays a large amount of useful information on any WC, or operator of the SMEFT or the LEFT specified by the user. For instance,
\begin{lstlisting}[style=mathematica]
ObjectInfo[CHu]
\end{lstlisting}
displays a menu with information about the SMEFT WC $C_{\varphi u}$, including the EFT to which it belongs, the name of the associated WC, the dimension (2, 3, 4, 5 or 6) and type (0F, 2F or 4F), whether it corresponds to an Hermitian operator or not, the number of independent real parameters, the number of flavor indices and the list of elements, as shown in Fig.~\ref{fig:dict}. A clickable menu with information about the SMEFT and LEFT parameters can be loaded with \rut{SMEFTOperatorsMenu}, \rut{LEFTOperatorsMenu} and \rut{TotalOperatorsMenu}, while grid menus with all the SMEFT or LEFT parameters can be generated with \rut{SMEFTOperatorsGrid} and \rut{LEFTOperatorsGrid}.
These grids are interactive, and the definition of any operator appears o screen when dragging the mouse pointer on top (see~Fig.~\ref{fig:grid}). In addition, clicking on the corresponding operator creates a pop-up window with the same chart created by \rut{ObjectInfo}.
The \mathe notebook {\tt OperatorsGrid.nb} can be found in the main \dsix folder. This notebook already contains the result of using \rut{OperatorsGridSMEFT} and \rut{OperatorsGridLEFT}, and the two grid menus can be used right after opening the notebook, without any need to load \dsix. This can be useful as an {\it out of the box} visual reference on the SMEFT and the LEFT.

\bigskip

Concerning handy routines for handling WCs and expressions with WCs (such as amplitudes or cross-sections), we highlight \rut{D6Simplify}. This routine is used to simplify expressions involving SMEFT or LEFT parameters, by replacing all redundant WCs in terms of non-redundant ones and eliminating complex conjugates on real parameters. For instance,
\begin{lstlisting}[style=mathematica]
D6Simplify[2 m2 CHq1[3,2] CC[Gd[3,1]]]
\end{lstlisting}
returns {\tt 2 m2 CHq1[2,3]$^\ast$ Gd[3,1]$^\ast$}, where the hermiticity relation $[C_{\varphi q}^{(1)}]_{32} = [C_{\varphi q}^{(1)}]_{23}^\ast$ for the SMEFT object $C_{\varphi q}^{(1)}$ has been used in order to express the result in terms of the independent parameter $[C_{\varphi q}^{(1)}]_{23}$.
As already mentioned, the function \rut{NIndependnet} returns the number of independent real parameters in a given parameter.
Finally, the routines \rut{ToSymmetric}, \rut{ToSymmetricSingle}, \rut{ToIndependent} and \rut{ToIndependentSingle} can be used to transform WCs to the symmetric and independent bases (see Appendix~\ref{ap:bases} for the definition of these bases). The routine \rut{CheckAndSymmetrize} also checks whether all hermiticity and antisymmetry conditions are satisfied in a given argument.

\section{Summary}
\label{mysum}

\dsix is a \mathe package for simbolic and numerical operations within the SMEFT and the LEFT, facilitating the treatment of these two effective theories in a systematic and complete manner.
 
Here we have presented \dsixv{2.0}, a new and improved version of \dsix.
This version features the complete one-loop evolution from a high-energy scale $\Lambda_{\rm UV}>\Lambda_{\rm EW}$ (where the physics is described by the SMEFT) down to a low-energy scale $\Lambda_{\rm IR}<\Lambda_{\rm EW}$ (where the physics is described by the LEFT). This includes complete one-loop RGE evolution and complete one-loop matching at the EW scale.
In addition, the new version contains a large number of improvements regarding notation and utilities, operational efficiency and simplicity, user interface, input and output, a set of reference tools for the SMEFT and the LEFT, and a complete \mathe documentation system.

\dsix is a project that can be extended with future improvements, including additional tools and functionalities. The final outcome of this endeavour will be a complete and powerful framework for the systematic exploration of new physics models using the language of Effective Field Theories.

\section*{Acknowledgements}
We thank Alejandro Celis for his decisive contributions to the development of \dsix as a whole, and his early involvement in the development of this second version. 
We are also grateful to Jorge Terol-Calvo, for testing an early version of \dsixv{2.0}.
This work has been partially funded by the DFG Scientific Network ``DixTools'', Project Number 406385113,
and by Generalitat Valenciana, Project SEJI/2018/033.
The work of J.F. was supported by the Cluster of Excellence `Precision Physics, Fundamental Interactions, and Structure of Matter' (PRISMA+ EXC 2118/1) funded by the German Research Foundation (DFG) within the German Excellence Strategy (Project ID 39083149).  
A.V. acknowledges financial support from the Spanish grants FPA2017-85216-P (MINECO/AEI/FEDER, UE) and FPA2017-90566-REDC (Red Consolider MultiDark), and from MINECO through the ``Ram\'on y Cajal'' contract RYC2018-025795-I. 
J.V. acknowledges funding from the European Union's Horizon 2020 research and innovation programme under the Marie Sklodowska-Curie grant agreement No 700525
`NIOBE', and from the Spanish MINECO through the ``Ram\'on y Cajal'' program RYC-2017-21870.

\appendix


\section{Standard Model Effective Field Theory}
\label{sec:SMEFT}

The SMEFT is the EFT obtained after extending the SM Lagrangian with all operators invariant under the $SU(3)_c \times SU(2)_L \times U(1)_Y$ gauge group up to an arbitrary dimension. The Lagrangian for the SMEFT can be written as 
\begin{equation} \label{eq:SMEFT}
\mathcal{L}_{\rm SMEFT} = \mathcal{L}_{\rm SM} + \sum_k C_k^{(5)} Q_k^{(5)}  + \sum_k C_k^{(6)} Q_k^{(6)} + \mathcal{O}\left( \frac{1}{\Lambda_{\rm UV}^3} \right) \, .
\end{equation}
The dimensionful Wilson coefficients $C_k^{(5)}$ and $C_k^{(6)}$ are implicitly suppressed by $1/\Lambda_{\rm UV}$ and $1/\Lambda_{\rm UV}^2$, respectively, where $\Lambda_{\rm UV}$ is the EFT cutoff scale, assumed to be much larger than the electroweak scale. The implementation of the SMEFT in \dsix~mainly follows the conventions used in~\Reff{Grzadkowski:2010es}.\footnote{
The reader should keep in mind that these conventions differ from those used in~\cite{Jenkins:2013zja,Jenkins:2013wua,Alonso:2013hga}. The differences appear in the normalization of $\lambda$ and $m$, the definition of the Yukawa matrices and the name of the gauge couplings. However, \dsixv{2.0} adopts the convention of \cite{Jenkins:2013zja,Jenkins:2013wua,Alonso:2013hga} of introducing the EFT cutoff scale into the definition of the WCs.
}
The SM Lagrangian is given by
\begin{align} 
\mathcal{L}_{\rm SM} &=-\frac{1}{4} G_{\mu \nu}^A G^{A \mu \nu} -\frac{1}{4} W_{\mu \nu}^I W^{I \mu \nu} -\frac{1}{4} B_{\mu \nu} B^{\mu \nu} + \left( D_\mu \varphi \right)^\dagger \left( D^\mu \varphi \right) + m^2 \varphi^\dagger \varphi -   \frac{\lambda}{2} \left( \varphi^\dagger \varphi \right)^2 \nonumber \\[1mm]
& \hspace{-4mm}
+ i \left( \bar \ell \slashed{D} \ell + \bar e \slashed{D} e + \bar q \slashed{D} q + \bar u \slashed{D} u + \bar d \slashed{D} d \right) - \left( \bar \ell \Gamma_e e \varphi + \bar q \Gamma_u u \widetilde \varphi + \bar q \Gamma_d d \varphi + \hc \right) +\mathcal{L}_\theta \, . \label{eq:SMLag}
\end{align}
Here $G_{\mu\nu}^A$ ($A=1\dots8$), $W_{\mu\nu}^I$ ($I=1\dots3$) and $B_{\mu\nu}$ denote, respectively, the $SU(3)_c$, $SU(2)_L$ and $U(1)_Y$ field-strength tensors. The fields $\ell$ and $q$ correspond to the lepton and quark $SU(2)_L$ doublets of the SM, while $e,u,d$ are the SM right-handed fields. The Higgs $SU(2)_L$ doublet is denoted by $\varphi$. The Yukawa couplings $\Gamma_{e,u,d}$ are $3 \times 3$ matrices in flavor space. Using appropriate field redefinitions, and without loss of generality, one can choose a particular flavor basis where $\Gamma_e$ and $\Gamma_d$ are diagonal and $\Gamma_u=V_{\rm CKM}\, \hat \Gamma_u$, with $\hat \Gamma_u$ diagonal and $V_{\rm CKM}$ denoting the CKM matrix. This is the so-called {\it down basis}, and it is the default basis choice for \dsix. Another basis choice that is also useful is the {\it up basis}, where $\Gamma_e$ and $\Gamma_u$ are diagonal and $\Gamma_d=V_{\rm CKM}^\dagger\, \hat \Gamma_d$ with $\hat \Gamma_d$ diagonal. Note, however, that these bases are not stable under RGE evolution. The covariant derivative is defined as
\begin{equation}
D_\mu = \partial_\mu + i g_s T^A G_\mu^A + i g T^I W_\mu^I + i g^\prime Y B_\mu \, ,
\end{equation}
where $\{g_s,g,g^\prime\}$ and $\{G,W,B\}$ are, respectively, the $SU(3)_c$, $SU(2)_L$ and
$U(1)_Y$ gauge couplings and gauge fields, and $T^A$ and $T^I$ are the corresponding gauge group generators in the appropriate representations. The hypercharge assignments for the matter fields are given in Table~\ref{hypercharge}.
The $\theta$ terms are given by
\begin{align}
\mathcal{L}_{\theta} =  \frac{\theta^{\prime} g^{\prime 2 }}{  32 \pi^2 }  \widetilde B_{\mu \nu}  B^{\mu \nu} + 
 \frac{\theta  g^{2}}{  32 \pi^2 }  \widetilde W_{\mu \nu}^{I}  W^{\mu \nu}_{I}  +  \frac{\theta_s g_s^{2}}{  32 \pi^2 }  \widetilde G_{\mu \nu}^{A}  G^{\mu \nu}_{A}   \,,
\end{align}
with the dual tensors defined as $\widetilde X = \frac{1}{2}\epsilon_{\mu \nu \rho \sigma} X^{\rho \sigma}$ (with $\epsilon_{0123}= +1$). There is only one operator of dimension five, the so-called Weinberg operator, 
\begin{align}
Q_{\ell \ell \vp \vp}=\left( \widetilde \vp^\dagger \ell \right)^T C \left( \widetilde \vp^\dagger \ell \right)\,,
\end{align}
with $C$ denoting the Dirac charge conjugation matrix. This operator gives a Majorana mass term for the neutrinos after spontaneous symmetry breaking~\cite{Weinberg:1979sa}. A non-redundant basis of dimension-six operators was defined in~\cite{Grzadkowski:2010es}, the so called \textit{Warsaw basis}. Table~\ref{tab:SMEFToperators} classifies the SMEFT operators in the Warsaw basis indicating the number of independent operators in each category. We list the Baryon-number-conserving operators in Tables~\ref{pbsot},~\ref{mixt} and \ref{fft}.  Barring flavor structure, these constitute a total of 59 operators, some of which are non-Hermitian, yielding in total 76 real coefficients.  Taking into account flavor indices, the Baryon-number-conserving dimension-six Lagrangian contains 1350 CP-even and 1149 CP-odd operators, for a total of 2499 Hermitian operators~\cite{Alonso:2013hga}.  The complete set of independent dimension-six Baryon number violating operators were identified in \cite{Abbott:1980zj}.  Barring flavor structure, there are only 4 Baryon-number-violating operators.  These are listed in Table~\ref{bvt}. 

\begin{table}[t]
\caption{Hypercharge assignments in the SMEFT. \label{hypercharge}}
\renewcommand{\arraystretch}{1.6}
\begin{center}
\begin{tabular}{|ccccccc|}
\hline 
Field & $\ell_L$ & $e_R$ & $q_L$ & $u_R$ & $d_R$ & $\vp$ \\
\hline
$Y$ & $-\frac{1}{2}$ & $-1$ & $\frac{1}{6}$ & $\frac{2}{3}$ & $-\frac{1}{3}$ & $\frac{1}{2}$ \\
\hline
\end{tabular}
\end{center}
\end{table}

\begin{table}
\renewcommand{\arraystretch}{1.3}
\caption{SMEFT operators in the Warsaw basis. The third column lists
  the number of operators in the category whereas the last column
  indicates whether they violate baryon ($B$) or lepton ($L$) numbers.\label{tab:SMEFToperators}}
\begin{center}
\begin{tabular}{|c|c|c|c|}
\hline
dim & class & $\#$ operators & quantum numbers \\
\hline
5 & Dimension-five & 1 & $\Delta L = 2$ \\
6 & $X^3$ & 4 & \\
6 & $\varphi^6$ & 1 & \\
6 & $\varphi^4 D^2$ & 2 & \\
6 & $X^2 \varphi^2$ & 8 & \\
6 & $\psi^2 \varphi^3$ & 3 & \\
6 & $\psi^2 X \varphi$ & 8 & \\
6 & $\psi^2 \varphi^2 D$ & 8 & \\
6 & $\left( \bar L L \right) \left( \bar L L \right)$ & 5 & \\
6 & $\left( \bar R R \right) \left( \bar R R \right)$ & 7 & \\
6 & $\left( \bar L L \right) \left( \bar R R \right)$ & 8 & \\
6 & $\left( \bar L R \right) \left( \bar L R \right)$ & 4 & \\
6 & $\left( \bar L R \right) \left( \bar R L \right)$ & 1 & \\
6 & Baryon-number-violating & 4 &  $\Delta B = \Delta L = 1$ \\
\hline
\end{tabular}
\end{center}
\end{table}

\begin{table} 
\renewcommand{\arraystretch}{1.2}
\caption{SMEFT purely bosonic operators. \label{pbsot}}
\begin{center}
\begin{tabular}{|c|c||c|c|}
\hline  
\multicolumn{2}{|c||}{$X^3$} & \multicolumn{2}{|c|}{$X^2 \varphi^2$} \\
\hline
$Q_G$  & $f^{ABC} G_\mu^{A \nu} G_\nu^{B \rho} G_\rho^{C \mu}$ & $Q_{\varphi G}$ & $\varphi^\dagger \varphi G_{\mu \nu}^A G^{A \mu \nu}$ \\
$Q_{\widetilde G}$ & $f^{ABC} \widetilde G_\mu^{A \nu} G_\nu^{B \rho} G_\rho^{C \mu}$ & $Q_{\varphi B}$ & $\varphi^\dagger \varphi B_{\mu \nu} B^{\mu \nu}$ \\
$Q_W$ & $\epsilon^{IJK} W_\mu^{I \nu} W_\nu^{J \rho} W_\rho^{K \mu}$ & $Q_{\varphi W}$  & $\varphi^\dagger \varphi W_{\mu \nu}^I W^{I \mu \nu}$ \\
$Q_{\widetilde W}$ & $\epsilon^{IJK} \widetilde W_\mu^{I \nu} W_\nu^{J \rho} W_\rho^{K \mu}$ &  $Q_{\varphi W B}$  & $\varphi^\dagger \tau^I \varphi W_{\mu \nu}^I B^{\mu \nu}$  \\
\cline{1-2}  
\multicolumn{2}{|c||}{$\myv{\varphi^6}$} & $Q_{\varphi \widetilde G}$   & $\varphi^\dagger \varphi \widetilde G_{\mu \nu}^A G^{A \mu \nu}$  \\
\cline{1-2}
 $Q_{\varphi}$ &$\myv{\left( \varphi^\dagger \varphi \right)^3}$ & $Q_{\varphi \widetilde B}$ & $\varphi^\dagger \varphi \widetilde B_{\mu \nu} B^{\mu \nu}$ \\
 \cline{1-2}
\multicolumn{2}{|c||}{$\varphi^4 D^2$} & $Q_{\varphi \widetilde W}$ &  $\varphi^\dagger \varphi \widetilde W_{\mu \nu}^I W^{I \mu \nu}$ \\
\cline{1-2}
$Q_{\varphi \Box}$ & $\myv{\left( \varphi^\dagger \varphi \right) \Box \left( \varphi^\dagger \varphi \right)}$ & $Q_{\varphi \widetilde W B}$ & $\varphi^\dagger \tau^I \varphi \widetilde W_{\mu \nu}^I B^{\mu \nu}$ \\
$Q_{\varphi D}$ & $\left( \varphi^\dagger D^\mu \varphi \right)^\ast \left( \varphi^\dagger D_\mu \varphi \right)$ &  &  \\
\hline
\end{tabular}
\end{center}
\end{table}

\begin{table}
\renewcommand{\arraystretch}{1.2}
\caption{SMEFT mixed operators involving bosons and fermions. \label{mixt}}
\begin{center}
\begin{tabular}{|c|c||c|c|}
\hline
\multicolumn{2}{|c||}{$\psi^2 \varphi^3$} & \multicolumn{2}{|c|}{$\psi^2 \varphi^2 D$} \\
\hline
$Q_{u \varphi}$  & $\left( \varphi^\dagger \varphi \right) \left( \bar q u \widetilde \varphi \right)$  & $Q_{\varphi \ell}^{(1)}$ & $\left( \varphi^\dagger i \Dlr_\mu \varphi \right) \left( \bar \ell \gamma^\mu \ell \right)$ \\
$Q_{d \varphi}$  & $\left( \varphi^\dagger \varphi \right) \left( \bar q d \varphi \right)$  & $Q_{\varphi \ell}^{(3)}$ & $\left( \varphi^\dagger i \DlrImu \varphi \right) \left( \bar \ell \tau^I \gamma^\mu \ell \right)$ \\
$Q_{e \varphi}$ & $\left( \varphi^\dagger \varphi \right) \left( \bar \ell e \varphi \right)$ & $Q_{\varphi e}$ & $\left( \varphi^\dagger i \Dlr_\mu \varphi \right) \left( \bar e \gamma^\mu e \right)$ \\  \cline{1-2}
\multicolumn{2}{|c||}{$\myv{\psi^2 X \varphi}$} & $Q_{\varphi q}^{(1)}$ & $\left( \varphi^\dagger i \Dlr_\mu \varphi \right) \left( \bar q \gamma^\mu q \right)$ \\
\cline{1-2} 
 $Q_{e W}$ &  ${\rule{0cm}{0.3cm}\left( \bar \ell \sigma^{\mu \nu} e \right) \tau^I \varphi W_{\mu \nu}^I}$ &  $Q_{\varphi q}^{(3)}$& $\left( \varphi^\dagger i \DlrImu \varphi \right) \left( \bar q \tau^I \gamma^\mu q \right)$ \\
  $Q_{e B}$ & $\left( \bar \ell \sigma^{\mu \nu} e \right) \varphi B_{\mu \nu}$  & $Q_{\varphi u}$ & $\left( \varphi^\dagger i \Dlr_\mu \varphi \right) \left( \bar u \gamma^\mu u \right)$ \\
  $Q_{u G}$  &  $\left( \bar q \sigma^{\mu \nu} T^A u \right) \widetilde \varphi G_{\mu \nu}^A$  &  $Q_{\varphi d}$  & $\left( \varphi^\dagger i \Dlr_\mu \varphi \right) \left( \bar d \gamma^\mu d \right)$ \\
$Q_{u W}$ & $\left( \bar q \sigma^{\mu \nu} u \right) \tau^I \widetilde \varphi W_{\mu \nu}^I$ & $Q_{\varphi u d}$ & $\left( \widetilde \varphi^\dagger i D_\mu \varphi \right) \left( \bar u \gamma^\mu d \right)$ \\[0.1cm]
$Q_{u B}$ & $\left( \bar q \sigma^{\mu \nu} u \right) \widetilde \varphi B_{\mu \nu}$ &  &  \\[0.1cm]
$Q_{d G}$ & $\left( \bar q \sigma^{\mu \nu} T^A d \right) \varphi G_{\mu \nu}^A$ &  &  \\[0.1cm]
$Q_{d W}$ & $\left( \bar q \sigma^{\mu \nu} d \right) \tau^I \varphi W_{\mu \nu}^I$ &  &  \\[0.1cm]
$Q_{d B}$ & $\left( \bar q \sigma^{\mu \nu} d \right) \varphi B_{\mu \nu}$ &  &  \\
\hline
\end{tabular}
\end{center}
\end{table}

\begin{table}
\renewcommand{\arraystretch}{1.2}
\caption{SMEFT purely fermionic operators which preserve Baryon number. \label{fft}}
\begin{center}
\begin{tabular}{|c|c||c|c|}
\hline
\multicolumn{2}{|c||}{$\left( \bar L L \right) \left( \bar L L \right)$} & \multicolumn{2}{|c|}{$\left( \bar L L \right) \left( \bar R R \right)$} \\
\hline
$Q_{\ell \ell}$ & $\left( \bar \ell \gamma_\mu \ell \right) \left( \bar \ell \gamma^\mu \ell \right)$ & $Q_{\ell e}$ & $\left( \bar \ell \gamma_\mu \ell \right) \left( \bar e \gamma^\mu e \right)$ \\
$Q_{q q}^{(1)}$ & $\left( \bar q \gamma_\mu q \right) \left( \bar q \gamma^\mu q \right)$ & $Q_{\ell u}$ & $\left( \bar \ell \gamma_\mu \ell \right) \left( \bar u \gamma^\mu u \right)$ \\
$Q_{q q}^{(3)}$ & $\left( \bar q \gamma_\mu \tau^I q \right) \left( \bar q \gamma^\mu \tau^I q \right)$ & $Q_{\ell d}$ & $\left( \bar \ell \gamma_\mu \ell \right) \left( \bar d \gamma^\mu d \right)$ \\
$Q_{\ell q}^{(1)}$ & $\left( \bar \ell \gamma_\mu \ell \right) \left( \bar q \gamma^\mu q \right)$ & $Q_{q e}$ & $\left( \bar q \gamma_\mu q \right) \left( \bar e \gamma^\mu e \right)$ \\
$Q_{\ell q}^{(3)}$ & $\left( \bar \ell \gamma_\mu \tau^I \ell \right) \left( \bar q \gamma^\mu \tau^I q \right)$ & $Q_{q u}^{(1)}$ & $\left( \bar q \gamma_\mu q \right) \left( \bar u \gamma^\mu u \right)$ \\
\cline{1-2}
\multicolumn{2}{|c||}{$\left( \bar R R \right) \left( \bar R R \right)$} & $Q_{q u}^{(8)}$ & $\left( \bar q \gamma_\mu T^A q \right) \left( \bar u \gamma^\mu T^A u \right)$ \\
\cline{1-2}
$Q_{ee}$ & $\left( \bar e \gamma_\mu e \right) \left( \bar e \gamma^\mu e \right)$ & $Q_{q d}^{(1)}$ & $\left( \bar q \gamma_\mu q \right) \left( \bar d \gamma^\mu d \right)$ \\
$Q_{uu}$ & $\left( \bar u \gamma_\mu u \right) \left( \bar u \gamma^\mu u \right)$ & $Q_{q d}^{(8)}$ & $\left( \bar q \gamma_\mu T^A q \right) \left( \bar d \gamma^\mu T^A d \right)$ \\
\cline{3-4}
$Q_{dd}$ & $\left( \bar d \gamma_\mu d \right) \left( \bar d \gamma^\mu d \right)$ & \multicolumn{2}{|c|}{$\myv{\left( \bar L R \right) \left( \bar R L \right)}$} \\
\cline{3-4}
$Q_{eu}$ & $\left( \bar e \gamma_\mu e \right) \left( \bar u \gamma^\mu u \right)$ & $Q_{\ell e d q}$ & $\myv{\left( \bar \ell^j e \right) \left( \bar d q^j \right)}$ \\
\cline{3-4}
$Q_{ed}$ & $\left( \bar e \gamma_\mu e \right) \left( \bar d \gamma^\mu d \right)$ & \multicolumn{2}{|c|}{$\myv{\left( \bar L R \right) \left( \bar L R \right)}$} \\
\cline{3-4}
$\myv{Q_{ud}^{(1)}}$ & $\left( \bar u \gamma_\mu u \right) \left( \bar d \gamma^\mu d \right)$ & $Q_{q u q d}^{(1)}$ & $\left( \bar q^j u \right) \epsilon_{jk} \left( \bar q^k d \right)$ \\
$Q_{ud}^{(8)}$ & $\left( \bar u \gamma_\mu T^A u \right) \left( \bar d \gamma^\mu T^A d \right)$ & $Q_{q u q d}^{(8)}$ & $\left( \bar q^j T^A u \right) \epsilon_{jk} \left( \bar q^k T^A d \right)$ \\
 & & $Q_{\ell e q u}^{(1)}$ & $\left( \bar \ell^j e \right) \epsilon_{jk} \left( \bar q^k u \right)$ \\
 & & $Q_{\ell e q u}^{(3)}$ & $\left( \bar \ell^j \sigma_{\mu \nu} e \right) \epsilon_{jk} \left( \bar q^k \sigma^{\mu \nu} u \right)$ \\
\hline
\end{tabular}
\end{center}
\end{table}

\begin{table}
\renewcommand{\arraystretch}{1.3}
\caption{SMEFT Baryon-number-violating operators.   \label{bvt}}
\begin{center}
\begin{tabular}{|c|c|}
\hline
\multicolumn{2}{|c|}{Baryon-number-violating} \\
\hline
$Q_{duq\ell}$ & $\left( d^T C u \right) \left( q^T C \ell \right)$ \\
$Q_{qque}$ & $\left( q^T C q \right) \left( u^T C e \right)$ \\
$Q_{qqq\ell}$ & $\epsilon_{il} \epsilon_{jk} \left( q_i^T C q_j \right) \left( q_k^T C \ell_l \right)$ \\
$Q_{duue}$ & $\left( d^T C u \right) \left( u^T C e \right)$ \\
\hline
\end{tabular}
\end{center}
\end{table}

The beta functions for the SMEFT WCs $C_i$ are defined as
\begin{equation}
\frac{dC_i}{d \log \mu}  \equiv \frac{1}{16 \pi^2} \, \beta_i \, .
\end{equation}
Here $\mu$ is the renormalization scale, and $\beta_i$ are the individual beta functions of each WC. The complete set of one-loop beta functions for the SM and dimension-six WCs were computed in \cite{Jenkins:2013zja,Jenkins:2013wua,Alonso:2013hga,Alonso:2014zka}.  The beta functions in these references neglect the contributions to the running of the dimension-six WCs from two insertions of the dimension-five Weinberg operator. Given the smallness of neutrino masses, it is natural to expect that the scale suppressing this operator is much larger than the one of the dimension-six operators, which justifies having neglected these contributions. The beta function for the Weinberg operator can be found in \cite{Antusch:2001ck}.
The complete set of one-loop SMEFT beta functions can be read off directly from \dsix with the command \rut{$\beta$[{\it parameter}]}.


\section{Low-Energy Effective Field Theory}
\label{sec:LEFT}

The LEFT is the EFT below the electroweak scale after integrating out
the Higgs boson, the massive $W^\pm$ and $Z$ gauge bosons and the top
quark from the SM particle content, as well as any BSM degrees of freedom at or above the EW scale. The resulting theory is invariant
under the $SU(3)_c \times U(1)_Q$ gauge group and contains $n_u = 2$
up-type quarks, $n_d = 3$ down-type quarks, $n_e = 3$ charged leptons
and $n_\nu = 3$ left-handed neutrinos. The LEFT Lagrangian is given by
\begin{equation} \label{eq:LEFT}
\mathcal{L}_{\rm LEFT} = \mathcal{L}_{\rm QCD+QED} + \sum_k L_k^{(3)} \Op_k^{(3)} + \sum_k L_k^{(5)} \Op_k^{(5)}  + \sum_k L_k^{(6)} \Op_k^{(6)} + \mathcal{O}\left( \frac{1}{\Lambda_{\rm EW}^3} \right) \, .
\end{equation}
The dimensionful Wilson coefficients $L_k^{(5)}$ and $L_k^{(6)}$ are implicitly suppressed by $1/\Lambda_{\rm EW}$ and $1/\Lambda_{\rm EW}^2$, respectively,
where $\Lambda_{\rm EW}$ is the electroweak scale.  A
non-redundant basis of dimension-three, -five and -six operators was introduced in~\cite{Jenkins:2017jig}, and this will be known in the following as the \textit{San Diego basis}. Table \ref{tab:LEFToperators} classifies the LEFT operators in the San
Diego basis indicating the number of independent operators in each
category. Barring flavor structure and Hermitian conjugation there are 96 independent operators. It can be shown that no linear combination of these operators vanish after
the application of the equations of motion, which makes them
completely independent operators. We list these operators in Tables~\ref{leftop1} -
\ref{leftop5}, omitting flavor (and $SU(3)_c$ indices in the last tables) to simplify the notation. The only operator present at dimension 3 is a Majorana mass term for
the left-handed neutrinos, shown in Table~\ref{leftop1}. There are two categories of dimension-five operators, $\left( \nu \nu
\right) X$ and $\left( \bar L R \right) X$, both listed in Table~\ref{leftop2}. While the former violates
Lepton number in two units, the latter preserves both lepton and
Baryon numbers. All the dimension-five LEFT operators are dipole operators. The remaining 89 independent operators arise at dimension 6. There are 2 purely gluonic operators, shown in Table~\ref{leftop3}, 56 4-fermion operators that conserve both Baryon and Lepton
numbers, shown in Table~\ref{leftop4}, and 31 4-fermion operators that violate Baryon and/or Lepton numbers, shown in Table~\ref{leftop5}.
Assuming that the SMEFT is the correct theory above the EW scale, all parameters of the LEFT can be fixed though a matching calculation at the EW scale. The complete set of matching conditions in the Warsaw and San Diego bases are known at tree-level~\cite{Jenkins:2017dyc} and one-loop~\cite{Dekens:2019ept} orders. They can be can be read off directly from \dsix with the command \rut{MatchEW[{\it parameter}]}.

The implementation of the LEFT in \dsix follows the same (or analogous)
conventions as for the SMEFT.~\footnote{The reader should keep in mind that these conventions differ from those used in~\cite{Jenkins:2017jig,Jenkins:2017dyc}. The differences appear in the definition of the fermion mass matrices and the name of the strong gauge coupling.} The QCD and QED Lagrangian is given by
\begin{align} 
\mathcal{L}_{\rm QCD+QED} &= -\frac{1}{4} G_{\mu \nu}^A G^{A \mu \nu} - \frac{1}{4} F_{\mu \nu} F^{\mu \nu} + \theta_{\rm QCD} \, \frac{g_s^2}{32 \, \pi^2} \widetilde G_{\mu \nu}^{A}  G^{\mu \nu}_{A} + + \theta_{\rm QED} \, \frac{e^2}{32 \, \pi^2} \widetilde F_{\mu \nu}  F^{\mu \nu} \nonumber \\
& + \sum_{\psi = u,d,e,\nu_L} \overline \psi \, i \slashed{D} \psi - \left[ \, \sum_{\psi = u,d,e} \overline \psi_L M_\psi \psi_R + \hc \right] \, \, . \label{eq:QCDQED}
\end{align}
The Dirac mass matrices $M_u$ and $M_{e,d}$ are, respectively, $2 \times 2$ and $3 \times 3$ matrices in
flavor space and we will omit flavor indices whenever possible. The
absence of a Dirac mass matrix for the neutrinos is due to the fact
that right-handed neutrinos are not included in the LEFT. The
covariant derivative is defined as
\begin{equation}
D_\mu = \partial_\mu + i g_s T^A G_\mu^A + i e Q A_\mu \, ,
\end{equation}
where $g_s$ and $e$ are the $SU(3)_c$ and $U(1)_Q$ gauge couplings,
respectively, and $T^A$ ($A=1\dots8$) are the Gell-Mann matrices. The
gauge field tensors are defined as usual,
\begin{align}
G_{\mu \nu}^A & = \partial_\mu G_\nu^A - \partial_\nu G_\mu^A - g_s f^{ABC} G_\mu^B G_\nu^C \, , \\
F_{\mu \nu} & = \partial_\mu F_\nu - \partial_\nu F_\mu \, ,
\end{align}
with covariant derivatives
\begin{align}
\left( D_\rho G_{\mu \nu} \right)^A & = \partial_\rho G_{\mu \nu}^A - g_s f^{ABC} G_\rho^B G_{\mu \nu}^C \, , \\
\left( D_\rho F_{\mu \nu} \right) & = \partial_\rho F_{\mu \nu} \, .
\end{align}
Finally, dual tensors are defined as $\widetilde X = \frac{1}{2}
\epsilon_{\mu \nu \rho \sigma} X^{\rho \sigma}$ (with $\epsilon_{0123}
= +1$).

\begin{table}
\renewcommand{\arraystretch}{1.6}
\caption{LEFT operators in the San Diego basis. The third column lists
  the number of operators in the category whereas the last column
  indicates whether they violate baryon ($B$) or lepton ($L$) numbers.
\label{tab:LEFToperators}}
\begin{center}
\begin{tabular}{|c|c|c|c|}
\hline
dim & class & $\#$ operators & quantum numbers \\
\hline
3 & $\nu \nu$ & 1 & $\Delta L = 2$ \\
\hline
5 & $\left( \nu \nu \right) X$ & 1 & $\Delta L = 2$ \\
5 & $\left( \bar L R \right) X$ & 5 & \\
\hline
6 & $X^3$ & 2 & \\
6 & $\left( \bar L L \right) \left( \bar L L \right)$ & 12 & \\
6 & $\left( \bar R R \right) \left( \bar R R \right)$ & 7 & \\
6 & $\left( \bar L L \right) \left( \bar R R \right)$ & 19 & \\
6 & $\left( \bar L R \right) \left( \bar L R \right)$ & 15 & \\
6 & $\left( \bar L R \right) \left( \bar R L \right)$ & 3 & \\
6 & $\Delta L = 4$ & 1 & $\Delta L = 4$ \\
6 & $\Delta L = 2$ & 14 & $\Delta L = 2$ \\
6 & $\Delta B = \Delta L = 1$ & 9 & $\Delta B = \Delta L = 1$ \\
6 & $\Delta B = - \Delta L = 1$ & 7 & $\Delta B = - \Delta L = 1$ \\
\hline
\end{tabular}
\end{center}
\end{table}

\begin{table}
\renewcommand{\arraystretch}{1.2}
\caption{LEFT dimension-three operator. \label{leftop1}}
\begin{center}
\begin{tabular}{|c|c|}
\hline
\multicolumn{2}{|c|}{$\nu \nu$} \\
\hline
$\Op_\nu$ & $\nu_L^T C \nu_L$ \\
\hline
\end{tabular}
\end{center}
\end{table}

\begin{table}
\renewcommand{\arraystretch}{1.6}
\caption{LEFT dimension-five operators. \label{leftop2}}
\begin{center}
\begin{tabular}{|c|c||c|c|}
\hline
\multicolumn{2}{|c||}{$\left( \nu \nu \right) X$} & \multicolumn{2}{|c|}{$\left( \bar L R \right) X$} \\
\hline
$\Op_{\nu \gamma}$ & $\left( \nu_L^T C \sigma^{\mu \nu} \nu_L \right) F_{\mu \nu}$ & $\Op_{e \gamma}$ & $\left( \bar e_L \sigma^{\mu \nu} e_R \right) F_{\mu \nu}$ \\
 & & $\Op_{u \gamma}$ & $\left( \bar u_L \sigma^{\mu \nu} u_R \right) F_{\mu \nu}$ \\
 & & $\Op_{d \gamma}$ & $\left( \bar d_L \sigma^{\mu \nu} d_R \right) F_{\mu \nu}$ \\
 & & $\Op_{u G}$ & $\left( \bar u_L \sigma^{\mu \nu} T^A u_R \right) G^A_{\mu \nu}$ \\
 & & $\Op_{d G}$ & $\left( \bar d_L \sigma^{\mu \nu} T^A d_R \right) G^A_{\mu \nu}$ \\
\hline
\end{tabular}
\end{center}
\end{table}

\begin{table}
\renewcommand{\arraystretch}{1.2}
\caption{LEFT purely gluonic operators. \label{leftop3}}
\begin{center}
\begin{tabular}{|c|c|}
\hline
\multicolumn{2}{|c|}{$X^3$} \\
\hline
$\Op_G$ & $f^{ABC} G_\mu^{A \nu} G_\nu^{B \rho} G_\rho^{C \mu}$ \\
$\Op_{\widetilde G}$ & $f^{ABC} \widetilde G_\mu^{A \nu} G_\nu^{B \rho} G_\rho^{C \mu}$ \\
\hline
\end{tabular}
\end{center}
\end{table}

{
\begin{table}
\renewcommand{\arraystretch}{1.2}
\caption{LEFT Baryon and Lepton number conserving dimension-six operators. \label{leftop4}}
\begin{center}
\begin{adjustbox}{width=1\textwidth,center=\textwidth}
\begin{tabular}{|c|c||c|c||c|c|}
\hline
\multicolumn{2}{|c||}{$\left( \bar L L \right) \left( \bar L L \right)$} & \multicolumn{2}{|c||}{$\left( \bar L L \right) \left( \bar R R \right)$} & \multicolumn{2}{|c|}{$\left( \bar L R \right) \left( \bar L R \right)$} \\
\hline
$\Op_{\nu \nu}^{V,LL}$ & $\left( \bar \nu_L \gamma_\mu \nu_L \right) \left( \bar \nu_L \gamma^\mu \nu_L \right)$ & $\Op_{\nu e}^{V,LR}$ & $\left( \bar \nu_L \gamma_\mu \nu_L \right) \left( \bar e_R \gamma^\mu e_R \right)$ & $\Op_{ee}^{S,RR}$ & $\left( \bar e_L e_R \right) \left( \bar e_L e_R \right)$ \\
$\Op_{ee}^{V,LL}$ & $\left( \bar e_L \gamma_\mu e_L \right) \left( \bar e_L \gamma^\mu e_L \right)$ & $\Op_{e e}^{V,LR}$ & $\left( \bar e_L \gamma_\mu e_L \right) \left( \bar e_R \gamma^\mu e_R \right)$ & $\Op_{eu}^{S,RR}$ & $\left( \bar e_L e_R \right) \left( \bar u_L u_R \right)$ \\
$\Op_{\nu e}^{V,LL}$ & $\left( \bar \nu_L \gamma_\mu \nu_L \right) \left( \bar e_L \gamma^\mu e_L \right)$ & $\Op_{\nu u}^{V,LR}$ & $\left( \bar \nu_L \gamma_\mu \nu_L \right) \left( \bar u_R \gamma^\mu u_R \right)$ & $\Op_{eu}^{T,RR}$ & $\left( \bar e_L \sigma_{\mu \nu} e_R \right) \left( \bar u_L \sigma^{\mu \nu} u_R \right)$ \\
$\Op_{\nu u}^{V,LL}$ & $\left( \bar \nu_L \gamma_\mu \nu_L \right) \left( \bar u_L \gamma^\mu u_L \right)$ & $\Op_{\nu d}^{V,LR}$ & $\left( \bar \nu_L \gamma_\mu \nu_L \right) \left( \bar d_R \gamma^\mu d_R \right)$ & $\Op_{ed}^{S,RR}$ & $\left( \bar e_L e_R \right) \left( \bar d_L d_R \right)$ \\
$\Op_{\nu d}^{V,LL}$ & $\left( \bar \nu_L \gamma_\mu \nu_L \right) \left( \bar d_L \gamma^\mu d_L \right)$ & $\Op_{e u}^{V,LR}$ & $\left( \bar e_L \gamma_\mu e_L \right) \left( \bar u_R \gamma^\mu u_R \right)$ & $\Op_{ed}^{T,RR}$ & $\left( \bar e_L \sigma_{\mu \nu} e_R \right) \left( \bar d_L \sigma^{\mu \nu} d_R \right)$ \\
$\Op_{e u}^{V,LL}$ & $\left( \bar e_L \gamma_\mu e_L \right) \left( \bar u_L \gamma^\mu u_L \right)$ & $\Op_{e d}^{V,LR}$ & $\left( \bar e_L \gamma_\mu e_L \right) \left( \bar d_R \gamma^\mu d_R \right)$ & $\Op_{\nu e d u}^{S,RR}$ & $\left( \bar \nu_L e_R \right) \left( \bar d_L u_R \right)$ \\
$\Op_{e d}^{V,LL}$ & $\left( \bar e_L \gamma_\mu e_L \right) \left( \bar d_L \gamma^\mu d_L \right)$ & $\Op_{u e}^{V,LR}$ & $\left( \bar u_L \gamma_\mu u_L \right) \left( \bar e_R \gamma^\mu e_R \right)$ & $\Op_{\nu e d u}^{T,RR}$ & $\left( \bar \nu_L \sigma_{\mu \nu} e_R \right) \left( \bar d_L \sigma^{\mu \nu} u_R \right)$ \\
$\Op_{\nu e d u}^{V,LL}$ & $\left( \bar \nu_L \gamma_\mu e_L \right) \left( \bar d_L \gamma^\mu u_L \right)$ & $\Op_{d e}^{V,LR}$ & $\left( \bar d_L \gamma_\mu d_L \right) \left( \bar e_R \gamma^\mu e_R \right)$ & $\Op_{u u}^{S1,RR}$ & $\left( \bar u_L u_R \right) \left( \bar u_L u_R \right)$ \\
$\Op_{u u}^{V,LL}$ & $\left( \bar u_L \gamma_\mu u_L \right) \left( \bar u_L \gamma^\mu u_L \right)$ & $\Op_{\nu e d u}^{V,LR}$ & $\left( \bar \nu_L \gamma_\mu e_L \right) \left( \bar d_R \gamma^\mu u_R \right)$ & $\Op_{u u}^{S8,RR}$ & $\left( \bar u_L T^A u_R \right) \left( \bar u_L T^A u_R \right)$ \\
$\Op_{d d}^{V,LL}$ & $\left( \bar d_L \gamma_\mu d_L \right) \left( \bar d_L \gamma^\mu d_L \right)$ & $\Op_{u u}^{V1,LR}$ & $\left( \bar u_L \gamma_\mu u_L \right) \left( \bar u_R \gamma^\mu u_R \right)$ & $\Op_{u d}^{S1,RR}$ & $\left( \bar u_L u_R \right) \left( \bar d_L d_R \right)$ \\
$\Op_{u d}^{V1,LL}$ & $\left( \bar u_L \gamma_\mu u_L \right) \left( \bar d_L \gamma^\mu d_L \right)$ & $\Op_{u u}^{V8,LR}$ & $\left( \bar u_L \gamma_\mu T^A u_L \right) \left( \bar u_R \gamma^\mu T^A u_R \right)$ & $\Op_{u d}^{S8,RR}$ & $\left( \bar u_L T^A u_R \right) \left( \bar d_L T^A d_R \right)$ \\
$\Op_{u d}^{V8,LL}$ & $\left( \bar u_L \gamma_\mu T^A u_L \right) \left( \bar d_L \gamma^\mu T^A d_L \right)$ & $\Op_{u d}^{V1,LR}$ & $\left( \bar u_L \gamma_\mu u_L \right) \left( \bar d_R \gamma^\mu d_R \right)$ & $\Op_{d d}^{S1,RR}$ & $\left( \bar d_L d_R \right) \left( \bar d_L d_R \right)$ \\
\cline{1-2}
\multicolumn{2}{|c||}{$\left( \bar R R \right) \left( \bar R R \right)$} & $\Op_{u d}^{V8,LR}$ & $\left( \bar u_L \gamma_\mu T^A u_L \right) \left( \bar d_R \gamma^\mu T^A d_R \right)$ & $\Op_{d d}^{S8,RR}$ & $\left( \bar d_L T^A d_R \right) \left( \bar d_L T^A d_R \right)$ \\
\cline{1-2}
$\Op_{e e}^{V,RR}$ & $\left( \bar e_R \gamma_\mu e_R \right) \left( \bar e_R \gamma^\mu e_R \right)$ & $\Op_{d u}^{V1,LR}$ & $\left( \bar d_L \gamma_\mu d_L \right) \left( \bar u_R \gamma^\mu u_R \right)$ & $\Op_{uddu}^{S1,RR}$ & $\left( \bar u_L d_R \right) \left( \bar d_L u_R \right)$ \\
$\Op_{e u}^{V,RR}$ & $\left( \bar e_R \gamma_\mu e_R \right) \left( \bar u_R \gamma^\mu u_R \right)$ & $\Op_{d u}^{V8,LR}$ & $\left( \bar d_L \gamma_\mu T^A d_L \right) \left( \bar u_R \gamma^\mu T^A u_R \right)$ & $\Op_{uddu}^{S8,RR}$ & $\left( \bar u_L T^A d_R \right) \left( \bar d_L T^A u_R \right)$ \\
\cline{5-6}
$\Op_{e d}^{V,RR}$ & $\left( \bar e_R \gamma_\mu e_R \right) \left( \bar d_R \gamma^\mu d_R \right)$ & $\Op_{d d}^{V1,LR}$ & $\left( \bar d_L \gamma_\mu d_L \right) \left( \bar d_R \gamma^\mu d_R \right)$ & \multicolumn{2}{|c|}{$\left( \bar L R \right) \left( \bar R L \right)$} \\
\cline{5-6}
$\Op_{u u}^{V,RR}$ & $\left( \bar u_R \gamma_\mu u_R \right) \left( \bar u_R \gamma^\mu u_R \right)$ & $\Op_{d d}^{V8,LR}$ & $\left( \bar d_L \gamma_\mu T^A d_L \right) \left( \bar d_R \gamma^\mu T^A d_R \right)$ & $\Op_{eu}^{S,RL}$ & $\left( \bar e_L e_R \right) \left( \bar u_R u_L \right)$ \\
$\Op_{d d}^{V,RR}$ & $\left( \bar d_R \gamma_\mu d_R \right) \left( \bar d_R \gamma^\mu d_R \right)$ & $\Op_{uddu}^{V1,LR}$ & $\left( \bar u_L \gamma_\mu d_L \right) \left( \bar d_R \gamma^\mu u_R \right)$ & $\Op_{ed}^{S,RL}$ & $\left( \bar e_L e_R \right) \left( \bar d_R d_L \right)$ \\
$\Op_{u d}^{V1,RR}$ & $\left( \bar u_R \gamma_\mu u_R \right) \left( \bar d_R \gamma^\mu d_R \right)$ & $\Op_{uddu}^{V8,LR}$ & $\left( \bar u_L \gamma_\mu T^A d_L \right) \left( \bar d_R \gamma^\mu T^A u_R \right)$ & $\Op_{\nu e d u}^{S,RL}$ & $\left( \bar \nu_L e_R \right) \left( \bar d_R u_L \right)$ \\
$\Op_{u d}^{V8,RR}$ & $\left( \bar u_R \gamma_\mu T^A u_R \right) \left( \bar d_R \gamma^\mu T^A d_R \right)$ & & & & \\
\hline
\end{tabular}
\end{adjustbox}
\end{center}
\end{table}
}

\begin{table}
\renewcommand{\arraystretch}{1.2}
\caption{LEFT Baryon and/or Lepton number violating dimension-six operators. We use $C$ to denote the Dirac charge conjugation matrix. \label{leftop5}}
\begin{center}
\small
\begin{tabular}{|c|c||c|c||c|c|}
\hline
\multicolumn{2}{|c||}{$\Delta L = 2$} & \multicolumn{2}{|c||}{$\Delta B = \Delta L = 1$} & \multicolumn{2}{|c|}{$\Delta B = - \Delta L = 1$} \\
\hline
$\Op_{\nu e}^{S,LL}$ & $\left( \nu_L^T C \nu_L \right) \left( \bar e_R e_L \right)$ & $\Op_{udd}^{S,LL}$ & $\left( u_L^T C d_L \right) \left( d_L^T C \nu_L \right)$ & $\Op_{ddd}^{S,LL}$ & $\left( d_L^T C d_L \right) \left( \bar e_R d_L \right)$ \\
$\Op_{\nu e}^{T,LL}$ & $\left( \nu_L^T C \sigma_{\mu \nu} \nu_L \right) \left( \bar e_R \sigma^{\mu \nu} e_L \right)$ & $\Op_{duu}^{S,LL}$ & $\left( d_L^T C u_L \right) \left( u_L^T C e_L \right)$ & $\Op_{udd}^{S,LR}$ & $\left( u_L^T C d_L \right) \left( \bar \nu_L d_R \right)$ \\
$\Op_{\nu e}^{S,LR}$ & $\left( \nu_L^T C \nu_L \right) \left( \bar e_L e_R \right)$ & $\Op_{uud}^{S,LR}$ & $\left( u_L^T C u_L \right) \left( d_R^T C e_R \right)$ & $\Op_{ddu}^{S,LR}$ & $\left( d_L^T C d_L \right) \left( \bar \nu_L u_R \right)$ \\
$\Op_{\nu u}^{S,LL}$ & $\left( \nu_L^T C \nu_L \right) \left( \bar u_R u_L \right)$ & $\Op_{duu}^{S,LR}$ & $\left( d_L^T C u_L \right) \left( u_R^T C e_R \right)$ & $\Op_{ddd}^{S,LR}$ & $\left( d_L^T C d_L \right) \left( \bar e_L d_R \right)$ \\
$\Op_{\nu u}^{T,LL}$ & $\left( \nu_L^T C \sigma_{\mu \nu} \nu_L \right) \left( \bar u_R \sigma^{\mu \nu} u_L \right)$ & $\Op_{uud}^{S,RL}$ & $\left( u_R^T C u_R \right) \left( d_L^T C e_L \right)$ & $\Op_{ddd}^{S,RL}$ & $\left( d_R^T C d_R \right) \left( \bar e_R d_L \right)$ \\
$\Op_{\nu u}^{S,LR}$ & $\left( \nu_L^T C \nu_L \right) \left( \bar u_L u_R \right)$ & $\Op_{duu}^{S,RL}$ & $\left( d_R^T C u_R \right) \left( u_L^T C e_L \right)$ & $\Op_{udd}^{S,RR}$ & $\left( u_R^T C d_R \right) \left( \bar \nu_L d_R \right)$ \\
$\Op_{\nu d}^{S,LL}$ & $\left( \nu_L^T C \nu_L \right) \left( \bar d_R d_L \right)$ & $\Op_{dud}^{S,RL}$ & $\left( d_R^T C u_R \right) \left( d_L^T C \nu_L \right)$ & $\Op_{ddd}^{S,RR}$ & $\left( d_R^T C d_R \right) \left( \bar e_L d_R \right)$ \\
\cline{5-6}
$\Op_{\nu d}^{T,LL}$ & $\left( \nu_L^T C \sigma_{\mu \nu} \nu_L \right) \left( \bar d_R \sigma^{\mu \nu} d_L \right)$ & $\Op_{ddu}^{S,RL}$ & $\left( d_R^T C d_R \right) \left( u_L^T C \nu_L \right)$ & \multicolumn{2}{|c|}{$\Delta L = 4$} \\
\cline{5-6}
$\Op_{\nu d}^{S,LR}$ & $\left( \nu_L^T C \nu_L \right) \left( \bar d_L d_R \right)$ & $\Op_{duu}^{S,RR}$ & $\left( d_R^T C u_R \right) \left( u_R^T C e_R \right)$ & $\Op_{\nu \nu}^{S,LL}$ & $\left( \nu_L^T C \nu_L \right) \left( \nu_L^T C \nu_L \right)$ \\
$\Op_{\nu e d u}^{S,LL}$ & $\left( \nu_L^T C e_L \right) \left( \bar d_R u_L \right)$ & & & & \\
$\Op_{\nu e d u}^{T,LL}$ & $\left( \nu_L^T C \sigma_{\mu \nu} e_L \right) \left( \bar d_R \sigma^{\mu \nu} u_L \right)$ & & & & \\
$\Op_{\nu e d u}^{S,LR}$ & $\left( \nu_L^T C e_L \right) \left( \bar d_L u_R \right)$ & & & & \\
$\Op_{\nu e d u}^{V,RL}$ & $\left( \nu_L^T C \gamma_\mu e_R \right) \left( \bar d_L \gamma^\mu u_L \right)$ & & & & \\
$\Op_{\nu e d u}^{V,RR}$ & $\left( \nu_L^T C \gamma_\mu e_R \right) \left( \bar d_R \gamma^\mu u_R \right)$ & & & & \\
\hline
\end{tabular}
\end{center}
\end{table}

The RGEs governing the renormalization scale evolution of the LEFT Wilson coefficients $L_i$ are given by
\begin{equation}
\frac{dL_i}{d \log \mu} =  \frac{1}{16 \pi^2} \, \beta_i \, .
\end{equation}
which define the LEFT beta functions $\beta_i$.
We use a notation completely analogous to that in the SMEFT.  The complete set of one-loop beta functions for the LEFT
has been computed in~\Reff{Jenkins:2017dyc}. 
They can be read off directly from DsixTools with the command \rut{$\beta$[{\it parameter}]}.


\section{SMEFT and LEFT parameters}
\label{ap:parameters}

In this Appendix we provide additional details about the variables used in \dsix. These can be useful to properly read and write some variables or apply some global dispatches and substitution rules in a \mathe session using \dsix. We also introduce the notation used in \dsix for the SMEFT and LEFT parameters.

\begin{table}
\small
\renewcommand{\arraystretch}{1.2}
\caption{Index symmetry categories used in \dsix.\label{tab:categories}}
\setlength{\tabcolsep}{10pt}
\begin{center}
\begin{tabular}{|cl|}
\hline
Category & \hspace{2cm} Meaning \\
\hline
0 & 0F scalar object \\
1 & 2F general $3 \times 3$ matrix \\
2 & 2F Hermitian matrix \\
3 & 2F symmetric matrix \\
4 & 2F antisymmetric matrix \\
5 & 4F general $3 \times 3 \times 3 \times 3$ object \\
6 & 4F two identical $\overline{\psi} \psi$ currents \\
7 & 4F two independent $\overline{\psi} \psi$ currents \\
8 & 4F two identical $\overline{\psi} \psi$ currents ($\psi$ singlet) \\
9 & 4F symmetric current $\times$ general current \\
10 & 4F antisymmetric current $\times$ general current \\
11 & 4F SMEFT special case $C_{qqql}$ \\
12 & 4F LEFT special case $L_{\nu \nu}^{S,LL}$ \\
13 & 4F LEFT special case $L_{ddd}^{S,LL/RR}$ \\
\hline
\end{tabular}
\end{center}
\end{table}

It is well known that some of the 2- and 4-fermion operators in the
SMEFT and the LEFT posess specific symmetries under the exchange of flavor
indices. 
For instance, the flavour components of the SMEFT operator $Q_{\vp e}$ form a Hermitian matrix, hence following the symmetry relation $[ Q_{\vp e}]_{ij} = [ Q_{\vp e}]_{ji}^\ast$, while the LEFT operator components of $\Op_{\nu \gamma}$ form an antisymmetric matrix, hence following the symmetry relation $[ \Op_{\nu \gamma}]_{ij} = -[ \Op_{\nu \gamma}]_{ji}$.
More complicated index symmetries exist for some of the 4-fermion operators. In all these cases, the number of independent operator components gets reduced, and thus the number of independent WCs. For example, the $C_{ee}$ 4-fermion WC does not contain $81$ ($=3^4$) independent complex WCs, but just $21$ real and $15$ imaginary independent components. It is convenient to restrict the number of parameters considered in SMEFT or LEFT calculations to just the independent
ones. In \dsix we have followed this approach, dropping redundant WCs in all internal calculations by transforming the user input into two minimal bases of operators: the {\it independent basis} and the {\it symmetric basis}. These bases, which are described in Appendix \ref{ap:bases}, have the same set of independent WCs, although with different numerical values. 
Since the number of independent WCs depends on the symmetry of the operators involved, it is sufficient to know the independent WCs for each index symmetry category of the operators in the SMEFT and LEFT.
The different categories are given in Table \ref{tab:categories}. We see that, apart from the operators belonging to categories 0, 1 and 5, all other operators have index symmetries. Furthermore, there are two dimension-six operators with special symmetries, not shared by any other operator, $Q_{qqql}$ and $\Op_{\nu \nu}^{S,LL}$, and two operators as only representatives of the last index symmetry category, $\Op_{ddd}^{S,LL}$ and $\Op_{ddd}^{S,RR}$. Similarly, the dimension-five SMEFT operator $Q_{\ell \ell \vp \vp}$ and the dimension-three LEFT neutrino mass matrix $M_\nu$ are the only symmetric matrices, while the dimension-five LEFT operator $\Op_{\nu \gamma}$ is the only antisymmetric matrix.

In Tables \ref{tab:nonredundant2F} and \ref{tab:nonredundant4F} we list the independent WCs contained in each category.  This, combined with Tables \ref{tab:SMEFTparameters} and \ref{tab:LEFTparameters}, completely allows the user to determine the position of a given parameter in the \rut{ParametersSMEFT} and \rut{ParametersLEFT} arrays. In any case, we remind the reader that the functions \rut{FindParametersSMEFT} and \rut{FindParametersLEFT} can also be used for this purpose.

{
\renewcommand{\arraystretch}{1.06}
\LTcapwidth=\textwidth
\begin{longtable}{|ccccc|}
\caption{\small Independent WCs in each 2F category. Numbers between curly brackets refer to the WC flavour indices. Elements in red denote real WCs. 
\label{tab:nonredundant2F}}\\
\hline
 & 1 & 2 & 3 & 4 \\
\hline
1 & \{1,1\} & \real{\{1,1\}} & \{1,1\} & \{1,2\} \\
2 & \{1,2\} & \{1,2\} & \{1,2\} & \{1,3\} \\
3 & \{1,3\} & \{1,3\} & \{1,3\} & \{2,3\} \\
4 & \{2,1\} & \real{\{2,2\}} & \{2,2\} &   \\
5 & \{2,2\} & \{2,3\} & \{2,3\} &   \\
6 & \{2,3\} & \real{\{3,3\}} & \{3,3\} &   \\
7 & \{3,1\} &   &   &   \\
8 & \{3,2\} &   &   &   \\
9 & \{3,3\} &   &   &   \\
\hline
\end{longtable}
}

\newpage
{
\footnotesize
\renewcommand{\arraystretch}{1.06}
\setlength{\tabcolsep}{3pt}
\LTcapwidth=\textwidth
\begin{longtable}{|cccccccccc|}
\caption{\small Independent WCs in each 4F category. Elements in red denote real WCs.
\label{tab:nonredundant4F}}\\
\hline
 & 5 & 6 & 7 & 8 & 9 & 10 & 11 & 12 & 13 \\
\hline
1 & \{1,1,1,1\} & \real{\{1,1,1,1\}} & \real{\{1,1,1,1\}} & \real{\{1,1,1,1\}} & \{1,1,1,1\} & \{1,2,1,1\} & \{1,1,1,1\} & \{1,1,2,2\} & \{1,2,1,1\} \\
2 & \{1,1,1,2\} & \{1,1,1,2\} & \{1,1,1,2\} & \{1,1,1,2\} & \{1,1,1,2\} & \{1,2,1,2\} & \{1,1,1,2\} & \{1,1,3,3\} & \{1,2,1,2\} \\
3 & \{1,1,1,3\} & \{1,1,1,3\} & \{1,1,1,3\} & \{1,1,1,3\} & \{1,1,1,3\} & \{1,2,1,3\} & \{1,1,1,3\} & \{2,2,3,3\} & \{1,2,1,3\} \\
4 & \{1,1,2,1\} & \real{\{1,1,2,2\}} & \real{\{1,1,2,2\}} & \real{\{1,1,2,2\}} & \{1,1,2,1\} & \{1,2,2,1\} & \{1,1,2,1\} & \{1,1,2,3\} & \{1,2,2,1\} \\
5 & \{1,1,2,2\} & \{1,1,2,3\} & \{1,1,2,3\} & \{1,1,2,3\} & \{1,1,2,2\} & \{1,2,2,2\} & \{1,1,2,2\} & \{1,2,2,3\} & \{1,2,2,2\} \\
6 & \{1,1,2,3\} & \real{\{1,1,3,3\}} & \real{\{1,1,3,3\}} & \real{\{1,1,3,3\}} & \{1,1,2,3\} & \{1,2,2,3\} & \{1,1,2,3\} & \{1,2,3,3\} & \{1,2,2,3\} \\
7 & \{1,1,3,1\} & \{1,2,1,2\} & \{1,2,1,1\} & \{1,2,1,2\} & \{1,1,3,1\} & \{1,2,3,1\} & \{1,1,3,1\} &   & \{1,2,3,1\} \\
8 & \{1,1,3,2\} & \{1,2,1,3\} & \{1,2,1,2\} & \{1,2,1,3\} & \{1,1,3,2\} & \{1,2,3,2\} & \{1,1,3,2\} &   & \{1,2,3,2\} \\
9 & \{1,1,3,3\} & \real{\{1,2,2,1\}} & \{1,2,1,3\} & \{1,2,2,2\} & \{1,1,3,3\} & \{1,2,3,3\} & \{1,1,3,3\} &   & \{1,2,3,3\} \\
10 & \{1,2,1,1\} & \{1,2,2,2\} & \{1,2,2,1\} & \{1,2,2,3\} & \{1,2,1,1\} & \{1,3,1,1\} & \{1,2,1,1\} &   & \{1,3,1,1\} \\
11 & \{1,2,1,2\} & \{1,2,2,3\} & \{1,2,2,2\} & \{1,2,3,2\} & \{1,2,1,2\} & \{1,3,1,2\} & \{1,2,1,2\} &   & \{1,3,1,2\} \\
12 & \{1,2,1,3\} & \{1,2,3,1\} & \{1,2,2,3\} & \{1,2,3,3\} & \{1,2,1,3\} & \{1,3,1,3\} & \{1,2,1,3\} &   & \{1,3,1,3\} \\
13 & \{1,2,2,1\} & \{1,2,3,2\} & \{1,2,3,1\} & \{1,3,1,3\} & \{1,2,2,1\} & \{1,3,2,1\} & \{1,2,2,1\} &   & \{1,3,2,1\} \\
14 & \{1,2,2,2\} & \{1,2,3,3\} & \{1,2,3,2\} & \{1,3,2,3\} & \{1,2,2,2\} & \{1,3,2,2\} & \{1,2,2,2\} &   & \{1,3,2,2\} \\
15 & \{1,2,2,3\} & \{1,3,1,3\} & \{1,2,3,3\} & \{1,3,3,3\} & \{1,2,2,3\} & \{1,3,2,3\} & \{1,2,2,3\} &   & \{1,3,2,3\} \\
16 & \{1,2,3,1\} & \{1,3,2,2\} & \{1,3,1,1\} & \real{\{2,2,2,2\}} & \{1,2,3,1\} & \{1,3,3,1\} & \{1,2,3,1\} &   & \{1,3,3,1\} \\
17 & \{1,2,3,2\} & \{1,3,2,3\} & \{1,3,1,2\} & \{2,2,2,3\} & \{1,2,3,2\} & \{1,3,3,2\} & \{1,2,3,2\} &   & \{1,3,3,2\} \\
18 & \{1,2,3,3\} & \real{\{1,3,3,1\}} & \{1,3,1,3\} & \real{\{2,2,3,3\}} & \{1,2,3,3\} & \{1,3,3,3\} & \{1,2,3,3\} &   & \{1,3,3,3\} \\
19 & \{1,3,1,1\} & \{1,3,3,2\} & \{1,3,2,1\} & \{2,3,2,3\} & \{1,3,1,1\} & \{2,3,1,1\} & \{1,3,1,1\} &   & \{2,3,1,2\} \\
20 & \{1,3,1,2\} & \{1,3,3,3\} & \{1,3,2,2\} & \{2,3,3,3\} & \{1,3,1,2\} & \{2,3,1,2\} & \{1,3,1,2\} &   & \{2,3,1,3\} \\
21 & \{1,3,1,3\} & \real{\{2,2,2,2\}} & \{1,3,2,3\} & \real{\{3,3,3,3\}} & \{1,3,1,3\} & \{2,3,1,3\} & \{1,3,1,3\} &   & \{2,3,2,2\} \\
22 & \{1,3,2,1\} & \{2,2,2,3\} & \{1,3,3,1\} &   & \{1,3,2,1\} & \{2,3,2,1\} & \{1,3,2,1\} &   & \{2,3,2,3\} \\
23 & \{1,3,2,2\} & \real{\{2,2,3,3\}} & \{1,3,3,2\} &   & \{1,3,2,2\} & \{2,3,2,2\} & \{1,3,2,2\} &   & \{2,3,3,2\} \\
24 & \{1,3,2,3\} & \{2,3,2,3\} & \{1,3,3,3\} &   & \{1,3,2,3\} & \{2,3,2,3\} & \{1,3,2,3\} &   & \{2,3,3,3\} \\
25 & \{1,3,3,1\} & \real{\{2,3,3,2\}} & \real{\{2,2,1,1\}} &   & \{1,3,3,1\} & \{2,3,3,1\} & \{1,3,3,1\} &   &   \\
26 & \{1,3,3,2\} & \{2,3,3,3\} & \{2,2,1,2\} &   & \{1,3,3,2\} & \{2,3,3,2\} & \{1,3,3,2\} &   &   \\
27 & \{1,3,3,3\} & \real{\{3,3,3,3\}} & \{2,2,1,3\} &   & \{1,3,3,3\} & \{2,3,3,3\} & \{1,3,3,3\} &   &   \\
28 & \{2,1,1,1\} &   & \real{\{2,2,2,2\}} &   & \{2,2,1,1\} &   & \{2,1,2,1\} &   &   \\
29 & \{2,1,1,2\} &   & \{2,2,2,3\} &   & \{2,2,1,2\} &   & \{2,1,2,2\} &   &   \\
30 & \{2,1,1,3\} &   & \real{\{2,2,3,3\}} &   & \{2,2,1,3\} &   & \{2,1,2,3\} &   &   \\
31 & \{2,1,2,1\} &   & \{2,3,1,1\} &   & \{2,2,2,1\} &   & \{2,1,3,1\} &   &   \\
32 & \{2,1,2,2\} &   & \{2,3,1,2\} &   & \{2,2,2,2\} &   & \{2,1,3,2\} &   &   \\
33 & \{2,1,2,3\} &   & \{2,3,1,3\} &   & \{2,2,2,3\} &   & \{2,1,3,3\} &   &   \\
34 & \{2,1,3,1\} &   & \{2,3,2,1\} &   & \{2,2,3,1\} &   & \{2,2,2,1\} &   &   \\
35 & \{2,1,3,2\} &   & \{2,3,2,2\} &   & \{2,2,3,2\} &   & \{2,2,2,2\} &   &   \\
36 & \{2,1,3,3\} &   & \{2,3,2,3\} &   & \{2,2,3,3\} &   & \{2,2,2,3\} &   &   \\
37 & \{2,2,1,1\} &   & \{2,3,3,1\} &   & \{2,3,1,1\} &   & \{2,2,3,1\} &   &   \\
38 & \{2,2,1,2\} &   & \{2,3,3,2\} &   & \{2,3,1,2\} &   & \{2,2,3,2\} &   &   \\
39 & \{2,2,1,3\} &   & \{2,3,3,3\} &   & \{2,3,1,3\} &   & \{2,2,3,3\} &   &   \\
40 & \{2,2,2,1\} &   & \real{\{3,3,1,1\}} &   & \{2,3,2,1\} &   & \{2,3,1,1\} &   &   \\
41 & \{2,2,2,2\} &   & \{3,3,1,2\} &   & \{2,3,2,2\} &   & \{2,3,1,2\} &   &   \\
42 & \{2,2,2,3\} &   & \{3,3,1,3\} &   & \{2,3,2,3\} &   & \{2,3,1,3\} &   &   \\
43 & \{2,2,3,1\} &   & \real{\{3,3,2,2\}} &   & \{2,3,3,1\} &   & \{2,3,2,1\} &   &   \\
44 & \{2,2,3,2\} &   & \{3,3,2,3\} &   & \{2,3,3,2\} &   & \{2,3,2,2\} &   &   \\
45 & \{2,2,3,3\} &   & \real{\{3,3,3,3\}} &   & \{2,3,3,3\} &   & \{2,3,2,3\} &   &   \\
46 & \{2,3,1,1\} &   &   &   & \{3,3,1,1\} &   & \{2,3,3,1\} &   &   \\
47 & \{2,3,1,2\} &   &   &   & \{3,3,1,2\} &   & \{2,3,3,2\} &   &   \\
48 & \{2,3,1,3\} &   &   &   & \{3,3,1,3\} &   & \{2,3,3,3\} &   &   \\
49 & \{2,3,2,1\} &   &   &   & \{3,3,2,1\} &   & \{3,1,3,1\} &   &   \\
50 & \{2,3,2,2\} &   &   &   & \{3,3,2,2\} &   & \{3,1,3,2\} &   &   \\
51 & \{2,3,2,3\} &   &   &   & \{3,3,2,3\} &   & \{3,1,3,3\} &   &   \\
52 & \{2,3,3,1\} &   &   &   & \{3,3,3,1\} &   & \{3,2,3,1\} &   &   \\
53 & \{2,3,3,2\} &   &   &   & \{3,3,3,2\} &   & \{3,2,3,2\} &   &   \\
54 & \{2,3,3,3\} &   &   &   & \{3,3,3,3\} &   & \{3,2,3,3\} &   &   \\
55 & \{3,1,1,1\} &   &   &   &   &   & \{3,3,3,1\} &   &   \\
56 & \{3,1,1,2\} &   &   &   &   &   & \{3,3,3,2\} &   &   \\
57 & \{3,1,1,3\} &   &   &   &   &   & \{3,3,3,3\} &   &   \\
58 & \{3,1,2,1\} &   &   &   &   &   &   &   &   \\
59 & \{3,1,2,2\} &   &   &   &   &   &   &   &   \\
60 & \{3,1,2,3\} &   &   &   &   &   &   &   &   \\
61 & \{3,1,3,1\} &   &   &   &   &   &   &   &   \\
62 & \{3,1,3,2\} &   &   &   &   &   &   &   &   \\
63 & \{3,1,3,3\} &   &   &   &   &   &   &   &   \\
64 & \{3,2,1,1\} &   &   &   &   &   &   &   &   \\
65 & \{3,2,1,2\} &   &   &   &   &   &   &   &   \\
66 & \{3,2,1,3\} &   &   &   &   &   &   &   &   \\
67 & \{3,2,2,1\} &   &   &   &   &   &   &   &   \\
68 & \{3,2,2,2\} &   &   &   &   &   &   &   &   \\
69 & \{3,2,2,3\} &   &   &   &   &   &   &   &   \\
70 & \{3,2,3,1\} &   &   &   &   &   &   &   &   \\
71 & \{3,2,3,2\} &   &   &   &   &   &   &   &   \\
72 & \{3,2,3,3\} &   &   &   &   &   &   &   &   \\
73 & \{3,3,1,1\} &   &   &   &   &   &   &   &   \\
74 & \{3,3,1,2\} &   &   &   &   &   &   &   &   \\
75 & \{3,3,1,3\} &   &   &   &   &   &   &   &   \\
76 & \{3,3,2,1\} &   &   &   &   &   &   &   &   \\
77 & \{3,3,2,2\} &   &   &   &   &   &   &   &   \\
78 & \{3,3,2,3\} &   &   &   &   &   &   &   &   \\
79 & \{3,3,3,1\} &   &   &   &   &   &   &   &   \\
80 & \{3,3,3,2\} &   &   &   &   &   &   &   &   \\
81 & \{3,3,3,3\} &   &   &   &   &   &   &   &   \\
\hline
\end{longtable}
}

\subsection{SMEFT parameters}
\label{ap:parameters-SMEFT}

\parskip2ex plus1ex 

Table \ref{tab:SMEFTparameters} provides a complete list of the SMEFT parameters used in \dsix. In addition to the SMEFT WCs, this includes the SM parameters (gauge couplings, Yukawa matrices and scalar and $\theta$ parameters). This table is particularly useful to identify the names given to the elements of 2- and 4-fermion WCs, as well as the corresponding beta functions, which can be readily obtained by evaluating {\tt \rut{$\beta$}[parameter]}. For instance, the beta function for the $g_s$ gauge coupling is obtained by evaluating \rut{$\beta$[gs]} and the beta function for the $[ C_{\ell q}^{(1)}]_{2233}$ WC is obtained with \rut{$\beta$[Clq1[2,2,3,3]]}.

\bigskip

{
\small
\LTcapwidth=\textwidth
\begin{longtable}{|cccccc|}
\caption{
SMEFT parameters. \emph{Position} denotes the position of the
  parameter (or parameters for 2- and 4-fermion objects) in the {\tt
    ParametersSMEFT} global array. \emph{Type} indicates the type of parameter (with nF
  standing for n-fermion) and \emph{Category}  denotes the index
  symmetry category of the coefficient, being relevant for 2- and
  4-fermion WCs.}
\label{tab:SMEFTparameters}\\
\hline
Position & Parameter(s) & \dsix name & Elements & Type & Category \\
\hline
1 & $g$ & {\tt g} & - & 0F & 0 \\
2 & $g^\prime$ & {\tt gp} & - & 0F & 0 \\
3 & $g_s$ & {\tt gs} & - & 0F & 0 \\
4 & $\lambda$ & {\tt $\lambda$} & - & 0F & 0 \\
5 & $m^2$ & {\tt m2} & - & 0F & 0 \\
6-14 & $\Gamma_u$ & {\tt MGu} & {\tt Gu[i,j]} & 2F & 1 \\
15-23 & $\Gamma_d$ & {\tt MGd} & {\tt Gd[i,j]} & 2F & 1 \\
24-32 & $\Gamma_e$ & {\tt MGe} & {\tt Ge[i,j]} & 2F & 1 \\
33 & $\theta$ & {\tt $\theta$} & - & 0F & 0 \\
34 & $\theta^\prime$ & {\tt $\theta$p} & - & 0F & 0 \\
35 & $\theta_s$ & {\tt $\theta$s} & - & 0F & 0  \\
36 & $C_G$ & {\tt CG} & - & 0F & 0 \\
37 & $C_{\widetilde G}$ & {\tt CGtilde} & - & 0F & 0 \\
38 & $C_W$ & {\tt CW} & - & 0F & 0 \\
39 & $C_{\widetilde W}$ & {\tt CWtilde} & - & 0F & 0 \\
40 & $C_\vp$ & {\tt CH} & - & 0F & 0 \\
41 & $C_{\vp\Box}$ & {\tt CHbox} & - & 0F & 0 \\
42 & $C_{\vp D}$ & {\tt CHD} & - & 0F & 0 \\
43 & $C_{\vp G}$ & {\tt CHG} & - & 0F & 0 \\
44 & $C_{\vp B}$ & {\tt CHB} & - & 0F & 0 \\
45 & $C_{\vp W}$ & {\tt CHW} & - & 0F & 0 \\
46 & $C_{\vp W B}$ & {\tt CHWB} & - & 0F & 0 \\
47 & $C_{\vp \widetilde G}$ & {\tt CHGtilde} & - & 0F & 0 \\
48 & $C_{\vp \widetilde B}$ & {\tt CHBtilde} & - & 0F & 0 \\
49 & $C_{\vp \widetilde W}$ & {\tt CHWtilde} & - & 0F & 0 \\
50 & $C_{\vp \widetilde W B}$ & {\tt CHWtildeB} & - & 0F & 0 \\
51-59 & $C_{u \vp}$ & {\tt MCuH} & {\tt CuH[i,j]} & 2F & 1 \\
60-68 & $C_{d \vp}$ & {\tt MCdH} & {\tt CdH[i,j]} & 2F & 1 \\
69-77 & $C_{e \vp}$ & {\tt MCeH} & {\tt CeH[i,j]} & 2F & 1 \\
78-86 & $C_{eW}$ & {\tt MCeW} & {\tt CeW[i,j]} & 2F & 1 \\
87-95 & $C_{eB}$ & {\tt MCeB} & {\tt CeB[i,j]} & 2F & 1 \\
96-104 & $C_{uG}$ & {\tt MCuG} & {\tt CuG[i,j]} & 2F & 1 \\
105-113 & $C_{uW}$ & {\tt MCuW} & {\tt CuW[i,j]} & 2F & 1 \\
114-122 & $C_{uB}$ & {\tt MCuB} & {\tt CuB[i,j]} & 2F & 1 \\
123-131 & $C_{dG}$ & {\tt MCdG} & {\tt CdG[i,j]} & 2F & 1 \\
132-140 & $C_{dW}$ & {\tt MCdW} & {\tt CdW[i,j]} & 2F & 1 \\
141-149 & $C_{dB}$ & {\tt MCdB} & {\tt CdB[i,j]} & 2F & 1 \\
150-155 & $C_{\vp \ell}^{(1)}$ & {\tt MCHl1} & {\tt CHl1[i,j]} & 2F & 2 \\
156-161 & $C_{\vp \ell}^{(3)}$ & {\tt MCHl3} & {\tt CHl3[i,j]} & 2F & 2 \\
162-167 & $C_{\vp e}$ & {\tt MCHe} & {\tt CHe[i,j]} & 2F & 2 \\
168-173 & $C_{\vp q}^{(1)}$ & {\tt MCHq1} & {\tt CHq1[i,j]} & 2F & 2 \\
174-179 & $C_{\vp q}^{(3)}$ & {\tt MCHq3} & {\tt CHq3[i,j]} & 2F & 2 \\
180-185 & $C_{\vp u}$ & {\tt MCHu} & {\tt CHu[i,j]} & 2F & 2 \\
186-191 & $C_{\vp d}$ & {\tt MCHd]} & {\tt CHd[i,j]} & 2F & 2 \\
192-200 & $C_{\vp u d}$ & {\tt MCHud} & {\tt CHud[i,j]} & 2F & 1 \\
201-227 & $C_{\ell \ell}$ & {\tt MCll} & {\tt Cll[i,j,k,l]} & 4F & 6 \\
228-254 & $C_{qq}^{(1)}$ & {\tt MCqq1} & {\tt Cqq1[i,j,k,l]} & 4F & 6 \\
255-281 & $C_{qq}^{(3)}$ & {\tt MCqq3} & {\tt Cqq3[i,j,k,l]} & 4F & 6 \\
282-326 & $C_{\ell q}^{(1)}$ & {\tt MClq1} & {\tt Clq1[i,j,k,l]} & 4F & 7 \\
327-371 & $C_{\ell q}^{(3)}$ & {\tt MClq3} & {\tt Clq3[i,j,k,l]} & 4F & 7 \\
372-392 & $C_{ee}$ & {\tt MCee} & {\tt Cee[i,j,k,l]} & 4F & 8 \\
393-419 & $C_{uu}$ & {\tt MCuu} & {\tt Cuu[i,j,k,l]} & 4F & 6 \\
420-446 & $C_{dd}$ & {\tt MCdd} & {\tt Cdd[i,j,k,l]} & 4F & 6 \\
447-491 & $C_{eu}$ & {\tt MCeu} & {\tt Ceu[i,j,k,l]} & 4F & 7 \\
492-536 & $C_{ed}$ & {\tt MCed} & {\tt Ced[i,j,k,l]} & 4F & 7 \\
537-581 & $C_{ud}^{(1)}$ & {\tt MCud1} & {\tt Cud1[i,j,k,l]} & 4F & 7 \\
582-626 & $C_{ud}^{(8)}$ & {\tt MCud8} & {\tt Cud8[i,j,k,l]} & 4F & 7 \\
627-671 & $C_{\ell e}$ & {\tt MCle} & {\tt Cle[i,j,k,l]} & 4F & 7 \\
672-716 & $C_{\ell u}$ & {\tt MClu} & {\tt Clu[i,j,k,l]} & 4F & 7 \\
717-761 & $C_{\ell d}$ & {\tt MCld} & {\tt Cld[i,j,k,l]} & 4F & 7 \\
762-806 & $C_{q e}$ & {\tt MCqe} & {\tt Cqe[i,j,k,l]} & 4F & 7 \\
807-851 & $C_{q u}^{(1)}$ & {\tt MCqu1} & {\tt Cqu1[i,j,k,l]} & 4F & 7 \\
852-896 & $C_{q u}^{(8)}$ & {\tt MCqu8} & {\tt Cqu8[i,j,k,l]} & 4F & 7 \\
897-941 & $C_{q d}^{(1)}$ & {\tt MCqd1} & {\tt Cqd1[i,j,k,l]} & 4F & 7 \\
942-986 & $C_{q d}^{(8)}$ & {\tt MCqd8} & {\tt Cqd8[i,j,k,l]} & 4F & 7 \\
987-1067 & $C_{\ell e d q}$ & {\tt MCledq} & {\tt Cledq[i,j,k,l]} & 4F & 5 \\
1068-1148 & $C_{quqd}^{(1)}$ & {\tt MCquqd1} & {\tt Cquqd1[i,j,k,l]} & 4F & 5 \\
1149-1229 & $C_{quqd}^{(8)}$ & {\tt MCquqd8} & {\tt Cquqd8[i,j,k,l]} & 4F & 5 \\
1230-1310 & $C_{\ell e q u}^{(1)}$ & {\tt MClequ1} & {\tt Clequ1[i,j,k,l]} & 4F & 5 \\
1311-1391 & $C_{\ell e q u}^{(3)}$ & {\tt MClequ3} & {\tt Clequ3[i,j,k,l]} & 4F & 5 \\
1392-1472 & $C_{duq\ell}$ & {\tt MCduql} & {\tt Cduql[i,j,k,l]} & 4F & 5 \\
1473-1526 & $C_{qque}$ & {\tt MCqque} & {\tt Cqque[i,j,k,l]} & 4F & 9 \\
1527-1583 & $C_{qqq\ell}$ & {\tt MCqqql} & {\tt Cqqql[i,j,k,l]} & 4F & 11 \\
1584-1664 & $C_{duue}$ & {\tt MCduue} & {\tt Cduue[i,j,k,l]} & 4F & 5 \\
1665-1670 & $C_{\ell\ell\vp\vp}$ & {\tt MCllHH} & {\tt CllHH[i,j]} & 2F & 3 \\
\hline
\end{longtable}
}

\subsection{LEFT parameters}
\label{ap:parameters-LEFT}

Table \ref{tab:LEFTparameters} provides a complete list of the LEFT parameters used in \dsix. In addition to the LEFT WCs, this includes the QCD and QED parameters (gauge couplings, fermion mass matrices and $\theta$ parameters). This table is particularly useful to identify the names given to the elements of 2- and 4-fermion WCs, as well as the corresponding beta functions, which can be readily obtained by evaluating {\tt \rut{$\beta$}[parameter]}. For instance, the beta function for the $e$ gauge coupling is obtained by evaluating \rut{$\beta$[eQED]} and the beta function for the $\left[ L_{dG}\right]_{22}$ WC is obtained with \rut{$\beta$[LdG[2,2]]}.
  
{
\small
\LTcapwidth=\textwidth
\begin{longtable}{|cccccc|}
\caption{
LEFT parameters. \emph{Position} denotes the position of the parameter
(or parameters for 2- and 4-fermion objects) in the {\tt
  ParametersLEFT} global array. \emph{Type} indicates the type
of parameter (with nF standing for n-fermion) and \emph{Category}
denotes the index symmetry category of the coefficient, being relevant
for 2- and 4-fermion WCs.}
\label{tab:LEFTparameters}\\
\hline
Position & Parameter(s) & \dsix name & Elements & Type & Category \\
\hline
1 & $g_s$ & {\tt gQCD} & - & 0F & 0 \\
2 & $e$ & {\tt eQED} & - & 0F & 0 \\
3 & $\theta_{\rm QCD}$ & {\tt $\theta$QCD} & - & 0F & 0 \\
4 & $\theta_{\rm QED}$ & {\tt $\theta$QED} & - & 0F & 0 \\
5-10 & $M_\nu$ & {\tt MM$\nu$} & {\tt M$\nu$[i,j]} & 2F & 3 \\
11-19 & $M_e$ & {\tt MMe} & {\tt Me[i,j]} & 2F & 1 \\
20-28 & $M_u$ & {\tt MMu} & {\tt Mu[i,j]} & 2F & 1 \\
29-37 & $M_d$ & {\tt MMd} & {\tt Md[i,j]} & 2F & 1 \\
38 & $L_G$ & {\tt LG} & - & 0F & 0 \\
39 & $L_{\widetilde G}$ & {\tt LGtilde} & - & 0F & 0 \\
40-42 & $L_{\nu \gamma}$ & {\tt ML$\nu \gamma$} & {\tt L$\nu \gamma$[i,j]} & 2F & 4 \\
43-51 & $L_{e \gamma}$ & {\tt MLe$\gamma$} & {\tt Le$\gamma$[i,j]} & 2F & 1 \\
52-60 & $L_{u \gamma}$ & {\tt MLu$\gamma$} & {\tt Lu$\gamma$[i,j]} & 2F & 1 \\
61-69 & $L_{d \gamma}$ & {\tt MLd$\gamma$} & {\tt Ld$\gamma$[i,j]} & 2F & 1 \\
70-78 & $L_{u G}$ & {\tt MLuG} & {\tt LuG[i,j]} & 2F & 1 \\
79-87 & $L_{d G}$ & {\tt MLdG} & {\tt LdG[i,j]} & 2F & 1 \\
88-108 & $L_{\nu \nu }^{V,LL}$ & {\tt ML$\nu\nu$VLL} & {\tt L$\nu\nu$VLL[i,j,k,l]} & 4F & 8 \\
109-129 & $L_{ee}^{V,LL}$ & {\tt MLeeVLL} & {\tt LeeVLL[i,j,k,l]} & 4F & 8 \\
130-210 & $L_{\nu e}^{V,LL}$ & {\tt ML$\nu$eVLL} & {\tt L$\nu$eVLL[i,j,k,l]} & 4F & 5 \\
211-291 & $L_{\nu u}^{V,LL}$ & {\tt ML$\nu$uVLL} & {\tt L$\nu$uVLL[i,j,k,l]} & 4F & 5 \\
292-372 & $L_{\nu d}^{V,LL}$ & {\tt ML$\nu$dVLL} & {\tt L$\nu$dVLL[i,j,k,l]} & 4F & 5 \\
373-453 & $L_{eu}^{V,LL}$ & {\tt MLeuVLL} & {\tt LeuVLL[i,j,k,l]} & 4F & 5 \\
454-534 & $L_{ed}^{V,LL}$ & {\tt MLedVLL} & {\tt LedVLL[i,j,k,l]} & 4F & 5 \\
535-615 & $L_{\nu edu}^{V,LL}$ & {\tt ML$\nu$eduVLL} & {\tt L$\nu$eduVLL[i,j,k,l]} & 4F & 5 \\
616-642 & $L_{uu}^{V,LL}$ & {\tt MLuuVLL} & {\tt LuuVLL[i,j,k,l]} & 4F & 6 \\
643-669 & $L_{dd}^{V,LL}$ & {\tt MLddVLL} & {\tt LddVLL[i,j,k,l]} & 4F & 6 \\
670-750 & $L_{ud}^{V1,LL}$ & {\tt MLudV1LL} & {\tt LudV1LL[i,j,k,l]} & 4F & 5 \\
751-831 & $L_{ud}^{V8,LL}$ & {\tt MLudV8LL} & {\tt LudV8LL[i,j,k,l]} & 4F & 5 \\
832-852 & $L_{ee}^{V,RR}$ & {\tt MLeeVRR} & {\tt LeeVRR[i,j,k,l]} & 4F & 8 \\
853-933 & $L_{eu}^{V,RR}$ & {\tt MLeuVRR} & {\tt LeuVRR[i,j,k,l]} & 4F & 5 \\
934-1014 & $L_{ed}^{V,RR}$ & {\tt MLedVRR} & {\tt LedVRR[i,j,k,l]} & 4F & 5 \\
1015-1041 & $L_{uu}^{V,RR}$ & {\tt MLuuVRR} & {\tt LuuVRR[i,j,k,l]} & 4F & 6 \\
1042-1068 & $L_{dd}^{V,RR}$ & {\tt MLddVRR} & {\tt LddVRR[i,j,k,l]} & 4F & 6 \\
1069-1149 & $L_{ud}^{V1,RR}$ & {\tt MLudV1RR} & {\tt LudV1RR[i,j,k,l]} & 4F & 5 \\
1150-1230 & $L_{ud}^{V8,RR}$ & {\tt MLudV8RR} & {\tt LudV8RR[i,j,k,l]} & 4F & 5 \\
1231-1311 & $L_{\nu e}^{V,LR}$ & {\tt ML$\nu$eVLR} & {\tt L$\nu$eVLR[i,j,k,l]} & 4F & 5 \\
1312-1392 & $L_{ee}^{V,LR}$ & {\tt MLeeVLR} & {\tt LeeVLR[i,j,k,l]} & 4F & 5 \\
1393-1473 & $L_{\nu u}^{V,LR}$ & {\tt ML$\nu$uVLR} & {\tt L$\nu$uVLR[i,j,k,l]} & 4F & 5 \\
1474-1554 & $L_{\nu d}^{V,LR}$ & {\tt ML$\nu$dVLR} & {\tt L$\nu$dVLR[i,j,k,l]} & 4F & 5 \\
1555-1635 & $L_{eu}^{V,LR}$ & {\tt MLeuVLR} & {\tt LeuVLR[i,j,k,l]} & 4F & 5 \\
1636-1716 & $L_{ed}^{V,LR}$ & {\tt MLedVLR} & {\tt LedVLR[i,j,k,l]} & 4F & 5 \\
1717-1797 & $L_{ue}^{V,LR}$ & {\tt MLueVLR} & {\tt LueVLR[i,j,k,l]} & 4F & 5 \\
1798-1878 & $L_{de}^{V,LR}$ & {\tt MLdeVLR} & {\tt LdeVLR[i,j,k,l]} & 4F & 5 \\
1879-1959 & $L_{\nu edu}^{V,LR}$ & {\tt ML$\nu$eduVLR} & {\tt L$\nu$eduVLR[i,j,k,l]} & 4F & 5 \\
1960-2040 & $L_{uu}^{V1,LR}$ & {\tt MLuuV1LR} & {\tt LuuV1LR[i,j,k,l]} & 4F & 5 \\
2041-2121 & $L_{uu}^{V8,LR}$ & {\tt MLuuV8LR} & {\tt LuuV8LR[i,j,k,l]} & 4F & 5 \\
2122-2202 & $L_{ud}^{V1,LR}$ & {\tt MLudV1LR} & {\tt LudV1LR[i,j,k,l]} & 4F & 5 \\
2203-2283 & $L_{ud}^{V8,LR}$ & {\tt MLudV8LR} & {\tt LudV8LR[i,j,k,l]} & 4F & 5 \\
2284-2364 & $L_{du}^{V1,LR}$ & {\tt MLduV1LR} & {\tt LduV1LR[i,j,k,l]} & 4F & 5 \\
2365-2445 & $L_{du}^{V8,LR}$ & {\tt MLduV8LR} & {\tt LduV8LR[i,j,k,l]} & 4F & 5 \\
2446-2526 & $L_{dd}^{V1,LR}$ & {\tt MLddV1LR} & {\tt LddV1LR[i,j,k,l]} & 4F & 5 \\
2527-2607 & $L_{dd}^{V8,LR}$ & {\tt MLddV8LR} & {\tt LddV8LR[i,j,k,l]} & 4F & 5 \\
2608-2688 & $L_{uddu}^{V1,LR}$ & {\tt MLudduV1LR} & {\tt LudduV1LR[i,j,k,l]} & 4F & 5 \\
2689-2769 & $L_{uddu}^{V8,LR}$ & {\tt MLudduV8LR} & {\tt LudduV8LR[i,j,k,l]} & 4F & 5 \\
2770-2796 & $L_{ee}^{S,RR}$ & {\tt MLeeSRR} & {\tt LeeSRR[i,j,k,l]} & 4F & 6 \\
2797-2877 & $L_{eu}^{S,RR}$ & {\tt MLeuSRR} & {\tt LeuSRR[i,j,k,l]} & 4F & 5 \\
2878-2958 & $L_{eu}^{T,RR}$ & {\tt MLeuTRR} & {\tt LeuTRR[i,j,k,l]} & 4F & 5 \\
2959-3039 & $L_{ed}^{S,RR}$ & {\tt MLedSRR} & {\tt LedSRR[i,j,k,l]} & 4F & 5 \\
3040-3120 & $L_{ed}^{T,RR}$ & {\tt MLedTRR} & {\tt LedTRR[i,j,k,l]} & 4F & 5 \\
3121-3201 & $L_{\nu edu}^{S,RR}$ & {\tt ML$\nu$eduSRR} & {\tt L$\nu$eduSRR[i,j,k,l]} & 4F & 5 \\
3202-3282 & $L_{\nu edu}^{T,RR}$ & {\tt ML$\nu$eduTRR} & {\tt L$\nu$eduTRR[i,j,k,l]} & 4F & 5 \\
3283-3309 & $L_{uu}^{S1,RR}$ & {\tt MLuuS1RR} & {\tt LuuS1RR[i,j,k,l]} & 4F & 6 \\
3310-3336 & $L_{uu}^{S8,RR}$ & {\tt MLuuS8RR} & {\tt LuuS8RR[i,j,k,l]} & 4F & 6 \\
3337-3417 & $L_{ud}^{S1,RR}$ & {\tt MLudS1RR} & {\tt LudS1RR[i,j,k,l]} & 4F & 5 \\
3418-3498 & $L_{ud}^{S8,RR}$ & {\tt MLudS8RR} & {\tt LudS8RR[i,j,k,l]} & 4F & 5 \\
3499-3525 & $L_{dd}^{S1,RR}$ & {\tt MLddS1RR} & {\tt LddS1RR[i,j,k,l]} & 4F & 6 \\
3526-3552 & $L_{dd}^{S8,RR}$ & {\tt MLddS8RR} & {\tt LddS8RR[i,j,k,l]} & 4F & 6 \\
3553-3633 & $L_{uddu}^{S1,RR}$ & {\tt MLudduS1RR} & {\tt LudduS1RR[i,j,k,l]} & 4F & 5 \\
3634-3714 & $L_{uddu}^{S8,RR}$ & {\tt MLudduS8RR} & {\tt LudduS8RR[i,j,k,l]} & 4F & 5 \\
3715-3795 & $L_{eu}^{S,RL}$ & {\tt MLeuSRL} & {\tt LeuSRL[i,j,k,l]} & 4F & 5 \\
3796-3876 & $L_{ed}^{S,RL}$ & {\tt MLedSRL} & {\tt LedSRL[i,j,k,l]} & 4F & 5 \\
3877-3957 & $L_{\nu edu}^{S,RL}$ & {\tt ML$\nu$eduSRL} & {\tt L$\nu$eduSRL[i,j,k,l]} & 4F & 5 \\
3958-3963 & $L_{\nu \nu }^{S,LL}$ & {\tt ML$\nu$$\nu$SLL} & {\tt L$\nu$$\nu$SLL[i,j,k,l]} & 4F & 12 \\
3964-4017 & $L_{\nu e}^{S,LL}$ & {\tt ML$\nu$eSLL} & {\tt L$\nu$eSLL[i,j,k,l]} & 4F & 9 \\
4018-4044 & $L_{\nu e}^{T,LL}$ & {\tt ML$\nu$eTLL} & {\tt L$\nu$eTLL[i,j,k,l]} & 4F & 10 \\
4045-4098 & $L_{\nu e}^{S,LR}$ & {\tt ML$\nu$eSLR} & {\tt L$\nu$eSLR[i,j,k,l]} & 4F & 9 \\
4099-4152 & $L_{\nu u}^{S,LL}$ & {\tt ML$\nu$uSLL} & {\tt L$\nu$uSLL[i,j,k,l]} & 4F & 9 \\
4153-4179 & $L_{\nu u}^{T,LL}$ & {\tt ML$\nu$uTLL} & {\tt L$\nu$uTLL[i,j,k,l]} & 4F & 10 \\
4180-4233 & $L_{\nu u}^{S,LR}$ & {\tt ML$\nu$uSLR} & {\tt L$\nu$uSLR[i,j,k,l]} & 4F & 9 \\
4234-4287 & $L_{\nu d}^{S,LL}$ & {\tt ML$\nu$dSLL} & {\tt L$\nu$dSLL[i,j,k,l]} & 4F & 9 \\
4288-4314 & $L_{\nu d}^{T,LL}$ & {\tt ML$\nu$dTLL} & {\tt L$\nu$dTLL[i,j,k,l]} & 4F & 10 \\
4315-4368 & $L_{\nu d}^{S,LR}$ & {\tt ML$\nu$dSLR} & {\tt L$\nu$dSLR[i,j,k,l]} & 4F & 9 \\
4369-4449 & $L_{\nu edu}^{S,LL}$ & {\tt ML$\nu$eduSLL} & {\tt L$\nu$eduSLL[i,j,k,l]} & 4F & 5 \\
4450-4530 & $L_{\nu edu}^{T,LL}$ & {\tt ML$\nu$eduTLL} & {\tt L$\nu$eduTLL[i,j,k,l]} & 4F & 5 \\
4531-4611 & $L_{\nu edu}^{S,LR}$ & {\tt ML$\nu$eduSLR} & {\tt L$\nu$eduSLR[i,j,k,l]} & 4F & 5 \\
4612-4692 & $L_{\nu edu}^{V,RL}$ & {\tt ML$\nu$eduVRL} & {\tt L$\nu$eduVRL[i,j,k,l]} & 4F & 5 \\
4693-4773 & $L_{\nu edu}^{V,RR}$ & {\tt ML$\nu$eduVRR} & {\tt L$\nu$eduVRR[i,j,k,l]} & 4F & 5 \\
4774-4854 & $L_{udd}^{S,LL}$ & {\tt MLuddSLL} & {\tt LuddSLL[i,j,k,l]} & 4F & 5 \\
4855-4935 & $L_{duu}^{S,LL}$ & {\tt MLduuSLL} & {\tt LduuSLL[i,j,k,l]} & 4F & 5 \\
4936-4962 & $L_{uud}^{S,LR}$ & {\tt MLuudSLR} & {\tt LuudSLR[i,j,k,l]} & 4F & 10 \\
4963-5043 & $L_{duu}^{S,LR}$ & {\tt MLduuSLR} & {\tt LduuSLR[i,j,k,l]} & 4F & 5 \\
5044-5070 & $L_{uud}^{S,RL}$ & {\tt MLuudSRL} & {\tt LuudSRL[i,j,k,l]} & 4F & 10 \\
5071-5151 & $L_{duu}^{S,RL}$ & {\tt MLduuSRL} & {\tt LduuSRL[i,j,k,l]} & 4F & 5 \\
5152-5232 & $L_{dud}^{S,RL}$ & {\tt MLdudSRL} & {\tt LdudSRL[i,j,k,l]} & 4F & 5 \\
5233-5259 & $L_{ddu}^{S,RL}$ & {\tt MLdduSRL} & {\tt LdduSRL[i,j,k,l]} & 4F & 10 \\
5260-5340 & $L_{duu}^{S,RR}$ & {\tt MLduuSRR} & {\tt LduuSRR[i,j,k,l]} & 4F & 5 \\
5341-5364 & $L_{ddd}^{S,LL}$ & {\tt MLdddSLL} & {\tt LdddSLL[i,j,k,l]} & 4F & 13 \\
5365-5445 & $L_{udd}^{S,LR}$ & {\tt MLuddSLR} & {\tt LuddSLR[i,j,k,l]} & 4F & 5 \\
5446-5472 & $L_{ddu}^{S,LR}$ & {\tt MLdduSLR} & {\tt LdduSLR[i,j,k,l]} & 4F & 10 \\
5473-5499 & $L_{ddd}^{S,LR}$ & {\tt MLdddSLR} & {\tt LdddSLR[i,j,k,l]} & 4F & 10 \\
5500-5526 & $L_{ddd}^{S,RL}$ & {\tt MLdddSRL} & {\tt LdddSRL[i,j,k,l]} & 4F & 10 \\
5527-5607 & $L_{udd}^{S,RR}$ & {\tt MLuddSRR} & {\tt LuddSRR[i,j,k,l]} & 4F & 5 \\
5608-5631 & $L_{ddd}^{S,RR}$ & {\tt MLdddSRR} & {\tt LdddSRR[i,j,k,l]} & 4F & 13 \\
\hline
\end{longtable}
}

\subsection{The symmetric and independent bases}
\label{ap:bases}

\dsix allows the user to introduce an arbitrary input for the WCs of the SMEFT in the Warsaw basis, and of the LEFT in the San Diego basis. 
In order to work only with independent parameters two different
operator bases are used in \dsix that drop all redundant WCs, and the user input and the results obtained from it can be expressed in terms of any of them. The first non-redundant basis, the {\it independent basis}, contains only the WCs with the flavour indices as listed in Tables \ref{tab:nonredundant2F} and \ref{tab:nonredundant4F} for each symmetry category, all other WCs being set to zero. In the second minimal basis, the {\it symmetric basis},  the redundancies in the WCs are removed by imposing that the latter follow the same symmetry relations as the corresponding operators. This is a convenient choice since the index symmetry of the operators is translated to the corresponding WCs. In order to simplify intermediate calculations done in this basis ({\it e.g.} in the RGEs) only the independent WCs listed for each category in Tables \ref{tab:nonredundant2F} and \ref{tab:nonredundant4F} are used. But unlike in the independent basis, the rest of WCs do not vanish but relate to the former following the same symmetry relations as the operators of the corresponding category. For instance, if we consider a 4-fermion operator with two identical $\bar{\psi}\psi$ currents ($\psi$ singlet) ({\it i.e.} either $Q_{ee}$ in the SMEFT, or $\Op_{ee}^{V,RR}$, $\Op_{ee}^{V,LL}$ and $\Op_{\nu\nu}^{V,LL}$ in the LEFT), which belongs to category 8 in Table \ref{tab:categories},
\begin{equation}\label{eq:4fop}
\sum_{prst} C_{prst} \, Q_{prst}  \, ,
\end{equation}
its WCs in the symmetric basis fulfill the relations $C_{stpr} = C_{prst}$ (because the two flavour currents are identical), $C_{rpts} = C_{prst}$ (due to hermiticity), and $C_{ptsr} = C_{prst}$ (as a consequence of the Fierz identity satisfied by the flavour components of the operator). Note that the sum in (\ref{eq:4fop}) runs over all possible values of the fermion flavour indices $(p,r,s,t)$. The same operator in the independent basis reads, however,
\begin{equation}
\sum_{\scriptsize \begin{array}{c} \{prst\}\in \rm \, cat.\, 8 \\ ({\rm real\,} \widetilde{C}) \end{array} } \widetilde{C}_{prst} \, Q_{prst}
\quad +
\sum_{\scriptsize \begin{array}{c} \{prst\}\in \rm \, cat.\, 8 \\ ({\rm complex\,} \widetilde{C}) \end{array} } \left( \,\widetilde{C}_{prst} \, Q_{prst} + \rm{h.c.} \right)
\, ,
\end{equation}
where now the sums comprise only the 21 (6 real and 15 complex)
independent components listed under category 8 of Table \ref{tab:nonredundant4F}, and all other $\widetilde{C}_{prst}$ vanish. 

It is straightforward to relate the WCs in the symmetric basis to those of the independent basis by using the symmetry relations satisfied by the operators. Let us provide an explicit example for illustration. Consider the contribution to the Lagrangian of the operator 
$\Op_{\nu\nu,prst}^{S,LL} = \left( \nu_{L,p}^T C \nu_{L,r} \right) \left( \nu_{L,s}^T C \nu_{L,t} \right)$ of the LEFT, which belongs to the symmetry category~12. Its flavour components are symmetric under the exchange of indices $p\leftrightarrow r$, $s\leftrightarrow t$ and $(p,r)\leftrightarrow (s,t)$, and further satisfy the Fierz identity $\Op_{prst} = - \Op_{ptsr} - \Op_{trsp}$. These relations reduce the number of independent components to just six. In the symmetric basis the contribution of this operator to the Lagrangian reads
\begin{equation}\label{eq:OpnunuSLL}
\sum_{prst} C_{prst} \, \Op_{\nu\nu,\,prst}^{S,LL}   \, ,
\end{equation}
where the $C_{prst}$ inherit the same index symmetries as those of the operator. Using those we can relate the 81 flavour components to the six independent ones chosen for category~12 (see Table \ref{tab:nonredundant4F}). In this way, (\ref{eq:OpnunuSLL}) reduces to
\begin{align}\label{eq:OpnunuSLL_ind}
& 3\,C_{1122} \, \Op_{\nu\nu,1122}^{S,LL} + 
6\,C_{1123} \, \Op_{\nu\nu,1123}^{S,LL} +
3\,C_{1133} \, \Op_{\nu\nu,1133}^{S,LL} \nn
\\ 
& + 24\,C_{1223} \, \Op_{\nu\nu,1223}^{S,LL} +
6\,C_{1233} \, \Op_{\nu\nu,1233}^{S,LL} +
3\,C_{2233} \, \Op_{\nu\nu,2233}^{S,LL}\, .
\end{align}
(\ref{eq:OpnunuSLL_ind}) matches the form of this operator in the independent basis, and thus allow us to read off the WCs in that basis in terms of the symmetric basis WCs:
\begin{align}
& \widetilde{C}_{1122} = 3\,C_{1122} \; , 
\widetilde{C}_{1123} = 6\,C_{1123} \; ,
\widetilde{C}_{1133} = 3\,C_{1133} \; ,
\\
&\widetilde{C}_{1223} = 24\,C_{1223} \; , 
\widetilde{C}_{1233} = 6\,C_{1233} \; ,
\widetilde{C}_{2233} = 3\,C_{2233} \, .
\end{align}
%


\section{Evolution matrix formalism}
\label{ap:evolution}

\dsixv{2.0} provides a new and much faster method of solving the RGE equations that relies on an semi-analytical solution of the RGE equations. To explain this method, we focus on the case where only dimension four and dimension six operators are present, and discuss the addition of dimension five operators at the end. The SMEFT and LEFT RGE equations can then be generically written as 
\begin{align}
\frac{dC_i^{\scriptscriptstyle (4)}(t)}{dt}&=\frac{1}{16\pi^2}\,\gamma_{ij}^{\scriptscriptstyle (4)}(C_k^{\scriptscriptstyle (4)},C_k^{\scriptscriptstyle (6)})\,C_j^{\scriptscriptstyle (4)}(t)\,,
\label{eq:RGEd4} \\[5pt]
\frac{d C_i^{\scriptscriptstyle (6)}(t)}{dt}&=\frac{1}{16\pi^2}\,\gamma_{ij}^{\scriptscriptstyle (6)}( C_k^{\scriptscriptstyle (4)})\, C_j^{\scriptscriptstyle (6)}(t)\,, 
\label{eq:RGEd6}
\end{align}
where $i,j,k$ span the number of EFT operators, $t\equiv\ln\mu$, and $\gamma$ is the anomalous dimension matrix (ADM). The superindices $(4)$ and $(6)$ denote, respectively, quantities associated to dimension four and six operators, and we have neglected contributions from $C_k^{\scriptscriptstyle (6)}$ in $\gamma^{\scriptscriptstyle (6)}$, since these correspond to higher orders in the EFT expansion. An analytical to solution to this system of coupled differential equations is not known, and one is generally forced to solve it numerically. Given the large number of equations involved, such numerical solution can be relatively slow. However, it is important to note that~\eqref{eq:RGEd6} still contains contributions that are higher order in the EFT expansion. Indeed, by noting that $C_k^{\scriptscriptstyle (6)}\sim\mathcal{O}(1/\Lambda^2)$, we can rewrite~\eqref{eq:RGEd4} as
\begin{align}\label{eq:RGEd4Red}
\frac{dC_i^{\scriptscriptstyle (4)}(t)}{dt}=\frac{1}{16\pi^2}\,\gamma_{ij}^{\scriptscriptstyle (4)}(C_k^{\scriptscriptstyle (4)})\,C_j^{\scriptscriptstyle (4)}(t)+\mathcal{O}(1/\Lambda^2)\,,
\end{align}
with $\gamma_{ij}^{\scriptscriptstyle (4)}(C_k^{\scriptscriptstyle (4)})\equiv\gamma_{ij}^{\scriptscriptstyle (4)}(C_k^{\scriptscriptstyle (4)},0)$. These equations correspond to the SM (or QCD and QED) RGE equations, and $\gamma_{ij}^{\scriptscriptstyle (4)}(C_k^{\scriptscriptstyle (4)})$ is known up to three loops~\cite{Bednyakov:2012rb,Bednyakov:2012en,Bednyakov:2013eba,Bednyakov:2014pia} and even up to five loops in QCD for the quark masses and QCD coupling~\cite{vanRitbergen:1997va, Vermaseren:1997fq,Baikov:2017ujl}. The numerical solution of this system of equations is much faster, given the reduced number of $C_k^{\scriptscriptstyle (4)}$ coefficients, and only needs to be performed once for a given set of experimental inputs. As a result, we get
\begin{align}
C_k^{\scriptscriptstyle (4)}(t)=\hat C_k^{\scriptscriptstyle (4)}(t)+\mathcal{O}(1/\Lambda^2)\,,
\end{align}
with $\hat C_k^{\scriptscriptstyle (4)}(t)$ being interpolating functions obtained from the numerical solution of~\eqref{eq:RGEd4Red}. Using this solution, we can rewrite~\eqref{eq:RGEd6} as
\begin{align}\label{eq:RGEd6Red}
\frac{dC_i^{\scriptscriptstyle (6)}(t)}{dt}&=\frac{1}{16\pi^2}\,\gamma_{ij}^{\scriptscriptstyle (6)}(\hat C_k^{\scriptscriptstyle (4)})\, C_j^{\scriptscriptstyle (6)}(t)+\mathcal{O}(1/\Lambda^4)\equiv \hat \gamma_{ij}^{\scriptscriptstyle (6)}(t)\, C_j^{\scriptscriptstyle (6)}(t)+\mathcal{O}(1/\Lambda^4)\,,
\end{align}
such that, up to corrections that are higher order in the EFT expansion, the ADM is just a function of $t$, completely fixed in terms of the interpolating functions $\hat C_k^{\scriptscriptstyle (4)}(t)$. Neglecting terms of $\mathcal{O}(1/\Lambda^2)$, the system of differential equations in~\eqref{eq:RGEd6Red} is solved by
\begin{align}\label{eq:RGEd6Sol}
C_i^{\scriptscriptstyle (6)}(t)=U^{\scriptscriptstyle (6)}_{ij}(t,t_0)\,C_j^{\scriptscriptstyle (6)}(t_0)\,,
\end{align}
where $t_0\equiv\ln\mu_0$, with $\mu_0$ being the input scale of the dimension-six WCs, and $U^{\scriptscriptstyle (6)}$ is an evolution matrix that is given in terms of $\hat\gamma^{\scriptscriptstyle (6)}(t)$ by\,\footnote{
In practice, it proves more convenient to determine the evolution matrix in~\eqref{eq:RGEd6Sol} by numerically solving~\eqref{eq:RGEd6Red} for a set of linearly independent $C_j^{\scriptscriptstyle (6)}(t_0)$ test inputs, rather than by using~\eqref{eq:Umatrix}.
}
\begin{align}\label{eq:Umatrix}
\begin{aligned}
U^{\scriptscriptstyle (6)}(t,t_0)&=\mathcal{T}\left\{\exp\left(\int_{t_0}^t \hat\gamma^{\scriptscriptstyle (6)}(\omega)\, d\omega\right)\right\}\\
&=\sum_{n=0}^\infty\int_{t_0}^t\int_{t_0}^{\omega_n}\int_{t_0}^{\omega_{n-1}}\dots\int_{t_0}^{\omega_2} \hat\gamma^{\scriptscriptstyle (6)}(\omega_1)\dots\hat\gamma^{\scriptscriptstyle (6)}(\omega_n)\, d\omega_1\dots d\omega_n\,.
\end{aligned}
\end{align}
Obtaining the evolution matrix is computationally expensive. However, since it is independent of the dimension-six input, it only needs to be determined once (for a given set of SM inputs). \dsixv{2.0} already contains a pre-computed evolution matrix for the inputs given in Table~\ref{tab:inputs}. Once the evolution matrix is known, the evaluation of~\eqref{eq:RGEd6Sol} is very fast. The solution for $C_i^{\scriptscriptstyle (6)}(t)$ in~\eqref{eq:RGEd6Sol} can then be plugged into the equations for the dimension-four WCs in~\eqref{eq:RGEd4}. These equations need to be solved numerically, but given the small number of equations, obtaining this numerical solution is considerably faster than solving the whole system.

Finally, we comment on the inclusion of dimension five operators, since these can potentially modify the method discussed here. The only dimension five operator in the SMEFT is the Weinberg operator. Since its WC is expected to be very small, given the smallness of the neutrino masses, we neglect its mixing to dimension-six SMEFT operators, which requires a double insertion of this operator. Once this contribution is neglected, the evolution matrix formalism can be trivially extended to include also the Weinberg operator. 
In the case of the LEFT, the presence of dimension-five dipole could be addressed by extending the above procedure order-by-order. The end result would be a numerical evolution matrix which takes into account the effect of double dipole insertions in the running of $C^{(6)}$. What we do is to produce the numerical evolution matrix neglecting double dipole insertions in the beta functions, and test the results of running with this evolution matrix against the exact results. We find that the agreement is numerically very accurate for all practical cases. However, the user should keep this in mind when considering applications with large contributions to dipole operators. In such situations it might be wise to compare the results of \rut{RGEsMethod}=3 with those obtained with \rut{RGEsMethod}=1 in a few cases. If significant effects from double dipole insertions are found, then running with \rut{RGEsMethod}=1 would be advised.

\bibliographystyle{JHEP}
\bibliography{references}

\end{document}